\documentclass[10pt,openany,a4paper]{article}
\usepackage{geometry,color,framed}
\usepackage{amsmath,amssymb,mathtools,slashed,xcolor,mdframed}
\usepackage[vcentermath,enableskew]{youngtab}
\usepackage{cancel, pifont,multirow}
\usepackage{tikz}
\usetikzlibrary{positioning}
\usetikzlibrary{decorations.text}
\usepackage{amstext}
\usepackage{graphicx}
\usepackage[section]{placeins}
\usepackage{epstopdf}
\usepackage{verbatim}
\usepackage{bm}
\usepackage{caption, subcaption, float,cite}
\usepackage[colorlinks=true, 
            linkcolor=blue, citecolor=blue,
            menucolor = red,
            urlcolor=blue,linktoc=all]{hyperref}
\makeatletter

\newcommand{\beq}{\begin{eqnarray}}
\newcommand{\eeq}{\end{eqnarray}}
\newcommand{\non}{\nonumber\\}

\newcommand{\mc}{\mathcal}
\newcommand{\mb}{\mathbb}

\newcommand{\ben}{\begin{enumerate}}
\newcommand{\bei}{\begin{itemize}}
\newcommand{\eni}{\end{itemize}}
\newcommand{\enn}{\end{enumerate}}
\newcommand{\ra}{\rightarrow}
\newcommand{\rr}{\right}
\newcommand{\lf}{\left}

\newcommand{\xdownarrow}[1]{%
  {\left\downarrow\vbox to #1{}\right.\kern-\nulldelimiterspace}
}

\def\l{{\lambda}}

\def\s{{\sigma}}

\usepackage{tikz}
\usetikzlibrary{matrix,shapes,arrows,positioning,chains}
\tikzset{
	 connector/.style={
        -latex,
        font=\scriptsize
    },
    rectangle connector/.style={
        connector,
        to path={(\tikztostart) -- ++(#1,0pt) \tikztonodes |- (\tikztotarget) },
        pos=0.5
    },
    rectangle connector/.default=-2cm,
    straight connector/.style={
        connector,
        to path=--(\tikztotarget) \tikztonodes
    }
}
\makeatother
\title{Hyperbolic Restricted Boltzmann Machine Neural Quantum State}
\author{H. L. Dao\footnote{espoirdujour1162@gmail.com}}
\date{\today}

\begin{document} 
\maketitle
\begin{abstract}
We construct the first type of  non-Euclidean non-autoregressive neural quantum state (NQS) in the form of the hyperbolic Restricted Boltzmann Machine (HRBM), which is studied in the variational Monte-Carlo (VMC) setting of the Quantum Sherrington-Kirkpatrick (QSK) model whose ground state exhibits volume-law entanglement. Across a 512-fold increase in the Hilbert space dimension corresponding to a system size increase from $N=14$ to $N=24$, HRBM NQS robustly outperforms its Euclidean version, the RBM NQS, in terms of better ground state energy optimization as well as lower R\'enyi-2 $S_2$ and von Neumann $S_{vN}$ absolute entanglement entropy reconstruction errors. More importantly, for all tested QSK system sizes, HRBM NQS demonstrates a superior expressivity in faithfully reproducing the entire entanglement spectrum of the QSK model from the top eigenvalues down to the tail end across 15 orders of magnitude, while RBM NQS consistently overestimates the sub-dominant modes. This work furnishes a proof-of-concept demonstrating that hyperbolic non-autoregressive NQS ans\"atze, thanks to the exponential volume of the hyperbolic geometry underlying their constructions, might be more natural at representing volume-law quantum systems than conventional Euclidean NQS. Furthermore, an interesting byproduct of this work is the polynomial scaling result of RBM-type NQS ans\"atze in the QSK volume-law system as the Hilbert space dimension increases exponentially.
\end{abstract}

\tableofcontents
\newpage
\section{Introduction}
Within quantum many-body systems, those possessing ground states that exhibit volume-law entanglement property such as quantum spin glasses, SYK and random matrix models (among others), are particularly interesting but difficult to study due to the exponential size of their Hilbert spaces. With the advent of neural quantum states (NQS) \cite{1606-rbm,1610-rbm, 1704-ann, 1709-ann, cnn-1807, nqs-2020, gcnn-2211, rnn_20, rnn-24, 2306-transformer, 2406-transformer} that are capable of efficiently representing many different types of quantum many-body systems\footnote{A particularly interesting study on the expressivity and generalization property of NQS in frustrated Heisenberg spin systems is the work \cite{nqs-2020}.}, these volume-law quantum systems represent a new challenging frontier. 
Many recent works, such as \cite{volume-2017}, \cite{volume-2019},  \cite{nn-volume-1}, \cite{nn-volume-2}, discussed and addressed the question of whether NQS can effectively model the ground states of these volume-law systems.  The results of the first few works \cite{volume-2017}, \cite{volume-2019} on this subject suggested that the Restricted Boltzmann Machine (RBM) NQS ans\"atze with polynomial numbers of parameters were capable of representing volume-law entanglement. In a later development, the work \cite{nn-volume-1} hypothesized that NQS (feedforward type) required an exponential scaling in order to represent volume-law systems such as the fermionic SYK model, while the follow-up work \cite{nn-volume-2} refuted this hypothesis by furnishing computational evidences showing that NQS did not require an exponential scaling to represent the volume-law SYK-like model if a suitable representation was chosen to take into account the fermionic sign structure of the fermionic states. In particular, the work \cite{nn-volume-2} considered two volume-law quantum systems, one being the SYK-type disordered fermionic system and the other being the quantum Sherrington-Kirkpatric (QSK) model \cite{qsk-ref-1}, \cite{qsk-ref-2} and showed that a suitably chosen representation of NQS only required a polynomial scaling to represent the volume-law ground states of these systems.
\\\\
Motivated by these recent developments, and by the results of our own works \cite{hnqs-1}, \cite{hnqs-2}, \cite{hnqs-3} in which the first few types of non-Euclidean NQS, in the form of hyperbolic Poincar\'e and Lorentz recurrent architectures, were constructed and applied in the transverse field Ising model (TFIM) and Heisenberg spin systems, we are interested in constructing new types of hyperbolic NQS for applications in volume-law quantum systems, since hyperbolic NQS ans\"atze have been shown to be able to outperform their Euclidean versions in prior works. In the field of Natural Language Processing (NLP) where hyperbolic neural networks \cite{ganea-18}, \cite{hyp-rep} were first introduced, the prime motivation for the constructions of these networks has to do with the exponential volume of hyperbolic space that allows for the enhanced expressivity\footnote{It is known that two-dimensional hyperbolic space can embed trees with arbitrarily low distortions \cite{sarkar}. Furthermore, it must be noted that the enhanced expressivity of hyperbolic neural networks in certain tasks compared to their Euclidean counterparts is purely thanks to the hyperbolic geometry underlying their constructions, since both types of neural networks have the exact same number of trainable parameters.} and superior performances of hyperbolic models compared to their Euclidean counterparts, especially in tasks where the data structure is tree-like or hierarchical \cite{hyp-rep}, \cite{hypnn-21}, \cite{hyp-survey}, or in graph-based tasks \cite{hyp-graph-19}, \cite{hyp-graph-chami}.
In the context of quantum-many body physics, in \cite{hnqs-1} and \cite{hnqs-2}, one-dimensional hyperbolic Lorentz and Poincar\'e RNN (recurrent neural network) and GRU (Gated Recurrent Unit) NQS were constructed and shown to outperform the Euclidean RNN and GRU NQS in $J_1J_2$ as well as $J_1J_2J_3$ Heisenberg quantum systems due to the hierarchical structure of the different degrees of nearest neighbor interactions, while in \cite{hnqs-3}, two-dimensional Lorentz RNN NQS was constructed and shown to outperform Euclidean RNN NQS in the two-dimensional transverse field Ising model at criticality where the physics can be described by a conformal field theory (CFT), which has an Anti-de-Sitter (AdS) dual\footnote{thanks to the AdS/CFT correspondance \cite{ads-cft}} whose spatial geometry is hyperbolic. 
\\\\
So far, these first developments of hyperbolic NQS ans\"atze in \cite{hnqs-1}, \cite{hnqs-2}, \cite{hnqs-3}  have involved only autoregressive recurrent architectures in quantum systems with area-law entanglement property. In volume-law quantum systems, the ground state requires an exponentially large Hilbert space capacity to represent dense entangling correlations, so the natural question that arises is, would some forms of hyperbolic NQS, either similar to those constructed recently in the works \cite{hnqs-1}, \cite{hnqs-2}, \cite{hnqs-3} or some entirely different architecture, provide a more natural variational wavefunction representation compared to Euclidean NQS, thanks to the hyperbolic geometry (with the associated exponential volume property) underlying all types of hyperbolic NQS ans\"atze? 
In this work, we attempt to address this question by presenting a proof-of-concept construction for a new type of hyperbolic NQS, the hyperbolic Restricted Boltzmann Machine (HRBM) NQS, which is the first hyperbolic non-autoregressive NQS introduced in the literature. The performance of HRBM is benchmarked against that of the regular or Euclidean RBM NQS in a Variational Monte Carlo (VMC) \cite{vmc_textbook} scaling study of the QSK model with system sizes $N$ ranging from 14 to 24 (across  a 512 fold increase in the Hilbert space dimension), where exact diagonalization is possible. The results of this work will demonstrate that HRBM NQS provides a definitive representational advantage over standard Euclidean RBM NQS across all tested system sizes. In particular, the hyperbolic latent space of HRBM provides a more robust geometric prior for representing volume-law  quantum spin-glass states. By efficiently accommodating hierarchical entanglement structures of the various QSK models under study, HRBM NQS achieves lower relative energy errors, lower energy variances, tighter energy bounds, and significantly higher entropic reconstruction fidelity as system size scales compared to RBM NQS.
\\\\
This paper is organized as follows. In Section \ref{hrbm}, we describe the mathematical construction of the hyperbolic RBM using the Lorentz model of hyperbolic space. In Section \ref{qsk_vmc}, we briefly describe the QSK model before describing in detail the VMC experiment settings (section \ref{vmc_settings}) and finally presenting the main results of this work, which includes a scaling study benchmarking the performance of the newly constructed HRBM NQS against that of the RBM NQS (see section \ref{scaling_results}), as well as a full entanglement spectrum study in section \ref{full_entanglement_results}. Section \ref{concl} summarizes the paper. In the Appendix, Section \ref{lor_math} includes the definitions of all relevant mathematical operations for the Lorentz model of hyperbolic space, Section \ref{exact_E_S} lists the exact ground state energies and $S_2$/$S_{vN}$ entropies of all QSK models considered in this work. These values serve as the exact references against which the performances of both HRBM and RBM NQS ans\"atze are benchmarked. Section \ref{scaling_seeds} includes the scaling study results for RBM and HRBM NQS at different realizations of QSK (corresponding to different QSK random seeds). Section \ref{ent_plots} includes the full entanglement spectrum plots corresponding to the exact wavefunction and the RBM/HRBM NQS at different QSK model sizes $N$. Finally, Section \ref{lmax_effects} includes the results of a comparison study at $N=24$ in which different HRBM ans\"atze with different hyperparameter $L_{max}$ are studied to choose the optimally performing ansatz.
\\\\
The Python codes to create and run the HRBM training as well as the trained model weights of the NQS constructed in this work are available at: \href{https://github.com/lorrespz/qsk_hrbm}{https://github.com/lorrespz/qsk\_hrbm}. In another departure from our earlier works on hyperbolic NQS which use \texttt{pytorch} and a standalone training pipeline, the HRBM NQS in this work is implemented with \texttt{jax}  \cite{jax2018github} and \texttt{flax.linen} \cite{flax2020github} which enable us to make use of \texttt{NetKet} \cite{netket-1}, \cite{netket-3} NQS training framework to perform the training (while the Euclidean RBM NQS already exists inside the \texttt{NetKet} library as a built-in function\footnote{callable as \texttt{netket.models.RBM} (\href{https://netket.readthedocs.io/en/latest/api/_generated/models/netket.models.RBM.html}{https://netket.readthedocs.io/en/latest/api/\_generated/models/netket.models.RBM.html})}).

\section{Hyperbolic RBM construction} \label{hrbm}
Before describing the construction of the hyperbolic RBM, we first recall the definition of the RBM. 
\\\\
An RBM consists of two layers of stochastic binary units, the $N$ visible units $v_i$ ($i=1,\ldots, N$) represent the observable data (such as physical spins in a quantum spin systems) and the $M$ hidden units $h_j$ ($j=1,\ldots, M)$ are the latent variables that capture correlations and features among the visible units. The network is restricted in the sense that there are no connections between units in the same layer (in other words, no intra-layer interactions). For a joint state configuration $(v, h)$, the network assigns an energy:
\beq
E(v, h) = -\sum_{i} a_i v_i - \sum_{j} b_j h_j - \sum_{i,j} v_i W_{ij} h_j
\eeq
where $a_i$ and $b_j$ are bias vectors for the visible and hidden layers, $W_{ij}$ is the weight matrix connecting visible unit $i$ to hidden unit $j$. The joint probability of a configuration is governed by the classical Boltzmann distribution:
\beq
P(v, h) = \frac{1}{Z} e^{-E(v, h)}
\eeq
where $Z = \sum_{v, h} e^{-E(v, h)}$ is the partition function. 
\\\\
To represent the spin-1/2 systems, the visible layer can be taken to be the physical spin variables ($\vec x=\s^z_1, \ldots, \s^z_N)$, so that RBM NQS is defined as follows:
\beq
\Psi_\text{RBM}(\vec x) = \sum_{\{h_j\}}\exp\left(\sum_i a_i x_i + \sum_j b_j h_j + \sum_{ij} W_{ij} h_j x_i\right)
\eeq
which, after tracing out the hidden units, becomes
\beq
\ln \Psi_{\text{RBM}}(\vec x) = \sum_{i=1}^N a_i x_i + \sum_{j=1}^M \ln\left[2 \cosh\left( b_j + \sum_{i=1}^N x_i W_{ij}\right) \right] \label{rbm-def}
\eeq
Note that the vectors $\vec a, \vec b$ and weight matrix $W$ can be either real (for a real NQS with just the amplitude part) or complex (for a complex NQS with both the amplitude and phase factor).
\\\\
To construct the hyperbolic RBM, in the case of a real NQS, the hidden weight vector $\vec b$ and the input vector $\vec x$ are mapped to a Lorentz hyperboloid via the  exponential map (given in Eq.\ref{eq-l-expmap}) while the visible weight vector $\vec a$ is kept Euclidean, alongside with the weight matrix $W$. 
The conversion of $b_j$ to a hyperbolic vector allows for a more expressive and flexible representation that is made possible by the exponential volume of the Lorentz hyperboloid.
\beq
\ln \Psi_{\text{HRBM}}(\vec x) = \sum_{i=1}^N a_i x_i + \sum_{j=1}^M \ln\left\{2 \cosh\lf[\log_{\mathbf{0}_\mc{L}}\left(\exp_{\mathbf{0}_\mc{L}}(b_j) \oplus_{\mc L} \sum_{i=1}^N  W_{ij} \otimes_{\mc L} \exp_{\mathbf{0}_\mc{L}} (x_i)\right) \right] \right\}
\label{hrbm-def}
\eeq
In Eq.\ref{hrbm-def} above, $\oplus_{\mc{L}}, \otimes_\mc{L}$ are the Lorentz addition and multiplication operations (Eq.\ref{lor-add} and Eq.\ref{lor-mult}), while 
 $\exp_{\mathbf{0}_\mc{L}}$, $\log_{\mathbf{0}_\mc{L}}$ are the exponential and logarithmic mappings between the Lorentz hyperboloid and its Euclidean tangent space. These hyperbolic Lorentz mathematical operations are defined in the Appendix \ref{lor_math}. This construction is similar to the hyperbolic recurrent NQS architectures of \cite{hnqs-1}-\cite{hnqs-3}, in the sense that only the bias vectors are embedded in hyperbolic space (either in the Poincar\'e or Lorentz model). The hyperbolic NQS constructions of \cite{hnqs-1}-\cite{hnqs-3}, are in turns based on the hyperbolic neural networks originally introduced in the work \cite{ganea-18} in the context of NLP. 
 \\\\
 For a complex NQS, a dual hyperbolic embedding is needed to take into account the real and imaginary components of the bias vector $\vec b$.
 \beq
 \ln \Psi_{\text{HRBM}}(\vec{x}) = \sum_{i=1}^N a_i x_i + \sum_{j=1}^M \ln \left[ 2 \cosh \left( \theta_{j, \text{re}}(\vec{x}) + i \, \theta_{j, \text{im}}(\vec{x}) \right) \right]
 \eeq
 where, for $\vec b=\vec b_\text{re} + i \vec b_\text{im}$ and $W = W_\text{re} + i W_\text{im}$,
 \beq
 \vec{\theta}_{\text{re}} &=& \log_{\mathbf{0}_\mc{L}} \left[\exp_{\mathbf{0}_\mc{L}}(\vec{b}_{\text{re}}) \oplus_\mc{L}  \lf(W_{\text{re}} \otimes_\mc{L} \exp_{\mathbf{0}_\mc{L}}(\vec{x})\rr) \right]
 \non
\vec{\theta}_{\text{im}} &=& \log_{\mathbf{0}_\mc{L}} \left[\exp_{\mathbf{0}_\mc{L}}(\vec{b}_{\text{im}}) \oplus_\mc{L}  \lf(W_{\text{im}} \otimes_\mc{L} \exp_{\mathbf{0}_\mc{L}}(\vec{x})\rr) \right] \label{hrbm-def}
\eeq
Given the defining equations of the RBM NQS (Eq.\ref{rbm-def}) and the HRBM NQS (Eq.\ref{hrbm-def}), we note that the two NQS architectures have the exact same number of learnable parameters, the complex-valued bias vectors $\vec a$, $\vec b$ and the weight matrix $W$. The major difference between HRBM and RBM is the underlying geometry of the manifold  on which mathematic operations such as matrix multiplication, bias addition and non-linear activation are carried out. As shown from our previous hyperbolic NQS constructions \cite{hnqs-1}, \cite{hnqs-2}, \cite{hnqs-3}, it is this exact geometry that is largely responsible for the difference in the expressive capacity of hyperbolic and Euclidean NQS ans\"atze,  since hyperbolic space with its exponential volume growth is capable of better representing hierarchical, tree-like systems, as opposed to  Euclidean space with its polynomial volume growth. On the other hand, to ensure numerical stability during the network training process, \cite{hnqs-2} introduces the spatial constraint hyperparameter, $L_{max}$\footnote{In the case of hyperbolic neural networks constructed using Lorentz hyperboloid,  $L_{max}$ restricts the spatial components $\vec x$ of the Lorentz vector $\mathbf{x} = (x_0, \vec{x})$ in order to prevent numerical instability due to open nature of the hyperboloid. The time component of the Lorentz vector, $x_0$, is constrained automatically once the spatial components are constrained, because of the defining equation of the Lorentz hyperboloid $\langle \mathbf{x}, \mathbf{x}\rangle_{\mc L} = -k$.}, for Lorentz-type hyperbolic NQS ans\"atze, which will be used for the HRBM constructed in this work. It must be noted that $L_{max}$ is not a trainable parameter but a hyperparameter (much like the learning rate) that should be chosen carefully to ensure optimal performances of the HRBM NQS. 
\section{QSK VMC Experiments} \label{qsk_vmc}
In this section, we describe the VMC experiment results involving the HRBM and RBM NQS applied to the QSK system, 
which describes a quantum spin glass model with random long-range interactions \cite{qsk-ref-1}, \cite{qsk-ref-2}. The Hamiltonian is given by
\beq
H = -\frac{1}{\sqrt{N}} \sum_{i < j} J_{ij} \sigma_i^z \sigma_j^z - \Gamma \sum_{i} \sigma_i^x
\eeq
where $\sigma_{x,z}$ are the Pauli $x$, $z$ matrices, $J_{ij}$ is the random spin glass coupling and $\Gamma$ is the transverse magnetic field along the $x$-axis. In this work, $\Gamma=-1$ is fixed throughout.
Note that the random all-to-all couplings $J_{ij}$ are normally distributed numbers with zero mean and they are responsible for the volume-law entanglement property in the ground state of the QSK model. At a fixed system size $N$, choosing a different random initialization seed for the model leads to a different set of $J_{ij}$ coefficients, hence a different realization of the QSK model at the same size $N$.

\subsection{VMC experiment settings} \label{vmc_settings}
The VMC experiments in this work are carried out for six different system sizes $N=14, 16, 18, 20, 22, 24$, spaning an increase of $2^{10}=512$ folds in the Hilbert size dimension, using two types of NQS ans\"atze: HRBM and RBM (see Table \ref{vmc_params} for the exact number of parameters). For each $N$, 12 different QSK random seeds are used corresponding to the 12 different sets of $J_{ij}$ couplings or 12 distinct QSK disorder realizations. For each fixed QSK seed, 5 different NQS seeds are used corresponding to different random initial configurations of the NQS training\footnote{As such, in total, for each QSK size $N$, 120 VMC experiment runs are performed corresponding to the two types of NQS, each using 5 different NQS seeds in 12 different realizations of the QSK model.}. The exact ground state energy and $S_2$/$S_{vN}$ entropies of all QSK models considered are listed in Tables \ref{exact_metrics_p1} and \ref{exact_metrics_p2} in Section \ref{exact_E_S}.
\begin{table}[!ht]
\centering
\begin{tabular}{c c c c c}
\hline\hline
& $N$& RBM & HRBM & \\
\hline\hline
& 14 & 224& 224\, ($L_{max}=10$)&  \\
& 16 & 288 & 288\, ($L_{max}=10$)&\\
& 18 & 360 & 360 \,($L_{max}=10$)& \\
& 20 & 440 & 440\, ($L_{max}=10$)&\\
& 22 & 528 & 528 \,($L_{max}=10$)& \\
& 24 & 624 & 624\, ($L_{max}=20$)& \\
\hline\hline
\end{tabular}
\caption{This table lists the different QSK system size $N$ considered and the corresponding number of parameters for the RBM/HRBM NQS ans\"atze. For a general $\alpha$ (the ratio of the hidden to visible units), the number of parameters for an RBM at QSK size $N$ is given by $N+ \alpha N + \alpha N^2$. When $\alpha=1$, the number of parameters is simply $(2+N)N$. Note that both RBM and HRBM ans\"atze use $\alpha=1$, and they have the exact same number of parameters at each $N$. As mentioned in the previous section, HRBM uses the additional hyperparameter $L_{max}$ that places a spatial constraint on the vectors in hyperbolic space to ensure numerical stability. For $N=14$ to $N=22$,  $L_{max}=10$ suffices to guarantee an optimal performance of HRBM, while for $N=24$, a larger $L_{max}=20$ is needed to ensure HRBM optimal performance. }\label{vmc_params}
\end{table}
\\\\
It is interesting to note that the number of RBM/HRBM parameters (listed in Table \ref{vmc_params}) are very small compared to the dimensions of the QSK Hilbert spaces (ranging from $2^{14}$ to $2^{24}$), but as the results show later, this very modest range of NQS parameters is sufficient to achieve the desired relative energy error that is smaller than $10^{-3}$.
For both RBM and HRBM NQS, the total number of training epochs is set to be 450, with an early stopping mechanism built in to stop the training if an early convergence is reached or if the mean energy fails to improve within 100 epochs. Similary, the learning rate, initially set at 0.05 is adjusted by a factor of 0.5 whenever the training runs into a plateau for 40 epochs. 
In all VMC experiments, we use the \texttt{NetKet} sampler \texttt{MetropolisLocal} with the number of samples being 1008 for $14\leq N\leq 22$ and $2016$ for $N=24$. As mentioned in the previous section and  in \cite{hnqs-1}, \cite{hnqs-2}, \cite{hnqs-3}, hyperbolic neural networks require a spatial constraint hyperparameter in order to perform stably. For all the HRBM NQS ans\"atze in this work, from $N=14$ to $N=22$, an $L_{max}=10$ suffices to guarantee an optimal performance of HRBM, while for $N=24$, a larger $L_{max}=20$ is needed to ensure HRBM optimal performance. This is because as $N$ expands, quantum state vectors extend further toward the boundary of the hyperbolic manifold, so increasing $L_{\max}$ at $N=24$ expands the accessible hyperbolic domain while keeping numerical overflow bounded. The choice of which $L_{max}$ to use for $N=24$ is made by checking the performances of different HRBM ans\"atze at different $L_{max}$ values in the range [8,10,12,15,20,25] and choosing the best-performing one in terms of energy and entropy error metrics (see Tables \ref{effects_lmax_energy} and \ref{effects_lmax_entropy} as well as Fig.\ref{n24_lmax_effects} in Section \ref{lmax_effects})\footnote{The results of Tables \ref{effects_lmax_energy} and \ref{effects_lmax_entropy} as well as Fig.\ref{n24_lmax_effects} show that as $L_{max}$ increases from 8, 10, 12, 15, to 20, there is a steady improvement in terms of the relative energy error, energy variance and entropy reconstruction errors, but as $L_{max}$ goes from 20 to 25, the improvement in terms of these metrics is minimal and although $L_{max}=25$ leads to a very slightly better performance of the HRBM NQS compared to $L_{max}=20$, choosing $L_{max}=20$ suffices for our computational purpose. Furthermore, it is interesting to note that at $N=24$, broadly speaking, HRBM ans\"atze with $L_{max} = 8, 10$ underperform RBM, HRBM with $L_{max}=12$ performs on par with RBM, while HRBM ans\"atze with $L_{max}=15, 20, 25$ outperform RBM. This means as the Hilbert space dimension increases, it is necessary to increase $L_{max}$ to enhance the representational capacity of the hyperbolic network while for smaller Hilbert space dimensions, keeping $L_{max}$ small is necessary to maintain numerical stability. }.
\\\\
Interestingly, despite the more complex construction of HRBM which requires hyperbolic mathematical operations (such as matrix exponentiation and Lorentzian additions $\oplus_{\mathcal{L}}$ among others), thanks to the new  \texttt{jax} implementation, there is no significant difference in the training time and computational overhead of HRBM compared to RBM, in direct constrast to our previous implementations that utilize \texttt{Tensorflow} and \texttt{pytorch}. For small system sizes up at $N=20$, the computations can be done on a modern machine (with around 8-16GB of RAM) with CPU alone in around an hour, while for system sizes $N=22, 24$, GPU computing is required to keep the training time within a few hours. 
\subsection{Scaling study results} \label{scaling_results}
\begin{figure}[!ht]
\centering
\includegraphics[width=.9\textwidth]{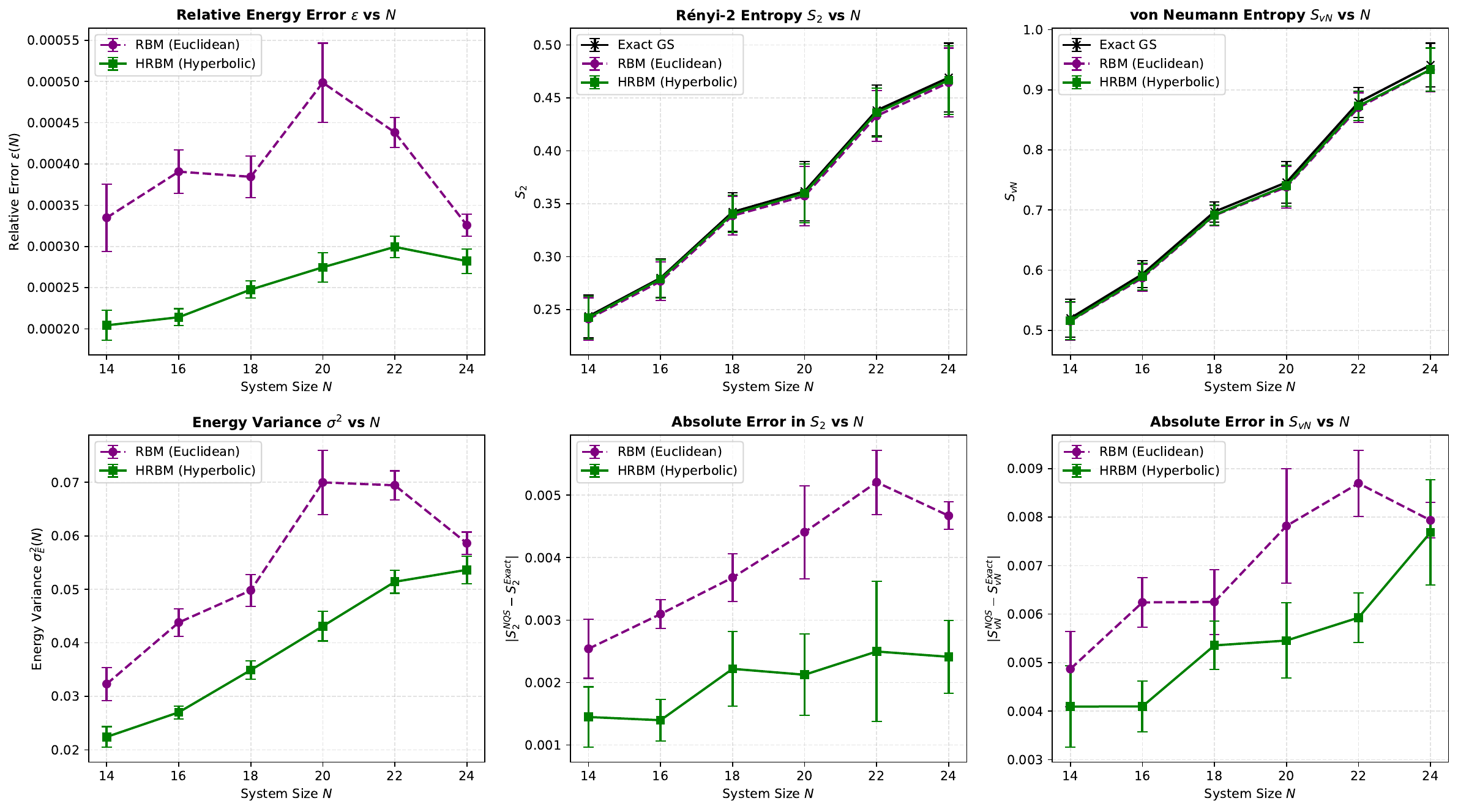}
\caption{The performance of HRBM NQS versus RBM NQS in the QSK system with increasing sizes from $N=14$ to $N=24$.
Top row, first subfigure from left: Relative energy error scaling ($\varepsilon= \langle |E_{\text{NQS}} - E_{\text{ED}}| / |E_{\text{ED}}| \rangle$ vs $N$) plot. This demonstrates how variational expressivity scales with Hilbert space dimension $2^N$. While RBM exhibits a non-monotonic optimization trajectory with a spike at $N=20$, HRBM displays a smooth, almost monotonic optimization trajectory across all QSK system sizes.
Top row, second subfigure from left: R\'enyi-2 $\langle S_2 \rangle$ entanglement entropy vs $N$ plot comparing disorder-averaged NQS $S_2$ entropy against the exact $S_2$ obtained by ED. 
Top row, third subfigure from left: von-Neumann $\langle S_{\text{vN}} \rangle$ entanglement entropy vs $N$ plot comparing disorder-averaged NQS entropies  against the exact $S_{vN}$ obtained by ED. This plot and the previous one show how each  ansatz tracks physical volume-law growth.
Bottom row, first subfigure from left: Energy variance scaling ($\sigma_E^2$ vs $N$) plot. This evaluates how closely each NQS architecture approaches the true exact Hamiltonian ground state wavefunction (where $\sigma_E^2 = 0$) as $N$ grows.
 Bottom row, second (third) subfigures from left: Absolute $S_2$ ($S_{vN}$) entropy reconstruction error $|S_2^{\text{NQS}} - S_{\text{Exact}}|$ ($|S_{vN}^{\text{NQS}} - S_{\text{Exact}}|$) plot. This highlights the exact entropy gap divergence as a function of system size.
 The error bars in all subfigures capture both the disorder variation stemming from the variance between different physical realizations of the random couplings $J_{ij}$ across the 12 QSK Hamiltonian seeds, as well as the NQS optimization stochastics originating from the variance in final network convergence due to the random weight initializations across the 5 different NQS seeds. The scaling results at individual QSK seeds corresponding to different QSK model realizations are included in Figs.\ref{scaling_s1111}-\ref{scaling_s9999} in Section \ref{scaling_seeds}) in the Appendix. }\label{main_scaling_res}
\end{figure}

To benchmark  HRBM NQS versus standard RBM NQS across system sizes $N$, we look at the various metrics that evaluate energy accuracy, state fidelity, and entanglement reconstruction such as the relative energy error $\varepsilon(N) = \langle |E_{\text{NQS}} - E_{\text{ED}}| / |E_{\text{ED}}| \rangle$, energy variance $\sigma_E^2(N)$, R\'enyi-2 $\langle S_2 \rangle$ and von Neumann $S_{vN}$ entropies\footnote{Note that the reduced density matrix $\rho=\text{Tr}(|\psi\rangle\langle \psi|)$, obtained by tracing out a subsystem of the quantum system under consideration, is used to define the set of R\'enyi entropies
\beq
S_n(\rho) = \frac{1}{1-n}\ln \text{Tr}(\rho^n) \eeq
When $n=1$, we obtain the von Neumann entropy $S_{vN} = - \text{Tr}\lf(\rho \log \rho\rr)$, while for $n=2$, we obtain the R\'enyi-2 entropy, $S_2 = -\ln(\text{Tr}\rho^2)$. In terms of the eigenvalues $\lambda_i$ of $\rho$,  $S_{vN} = -\sum_i \l_i \ln \l_i$ while the R\'enyi-2 entropy is defined as $S_2 = -\ln\sum_i \lambda_i^2$. We note that $S_2$ is almost entirely driven by the top $1\%$ of eigenmodes because squaring $\lambda_i$ suppresses small modes extremely aggressively (e.g. when $\lambda = 10^{-8}, \lambda^2 = 10^{-16}$). On the other hand, von Neumann entropy $S_{vN}$  provides better visibility into intermediate modes ($\lambda \in [10^{-2}, 10^{-4}]$) compared to $S_2$, but remains blind to deep tail modes ($\lambda < 10^{-8}$).},
and absolute $S_2$ ($S_{vN}$) entropy error $|S^\text{err}_{2}| = |S_2^{\text{NQS}} - S^\text{Exact}_2|$ ($S^\text{err}_{vN} = |S_{vN}^{\text{NQS}} - S^\text{Exact}_{vN}|$).
\\\\
 Our main results are shown in Tables \ref{main_rel_energy_error}, \ref{main_entropy} and graphically in Fig.\ref{main_scaling_res}.
\begin{table}[!ht]
\centering
\begin{tabular}{c c c c}
\hline\hline
$N$ & NQS & Mean relative energy error &  Mean variance \\
\hline\hline
14 & HRBM & $2.0427 \times 10^{-4}$ & $2.2411 \times 10^{-2}$ \\
-- & --- & $1.8116 \times 10^{-5}$ & $1.9192 \times 10^{-3}$ \\
-- & RBM & $3.3480 \times 10^{-4}$ & $3.2306 \times 10^{-2}$ \\
-- & --- & $4.0949 \times 10^{-5}$ & $3.0845 \times 10^{-3}$ \\
\hline
16 & HRBM & $2.1406 \times 10^{-4}$ & $2.7007 \times 10^{-2}$ \\
-- & --- & $1.0370 \times 10^{-5}$ & $1.2069 \times 10^{-3}$ \\
-- & RBM & $3.9070 \times 10^{-4}$ & $4.3801 \times 10^{-2}$ \\
-- & --- & $2.6566 \times 10^{-5}$ & $2.5479 \times 10^{-3}$ \\
\hline
18 & HRBM & $2.4763 \times 10^{-4}$ & $3.4905 \times 10^{-2}$ \\
-- & --- & $1.0327 \times 10^{-5}$ & $1.6723 \times 10^{-3}$ \\
-- & RBM & $3.8445 \times 10^{-4}$ & $4.9802 \times 10^{-2}$ \\
-- & --- & $2.5227 \times 10^{-5}$ & $2.9724 \times 10^{-3}$ \\
\hline
20 & HRBM & $2.7468 \times 10^{-4}$ & $4.3112 \times 10^{-2}$ \\
-- & --- & $1.7805 \times 10^{-5}$ & $2.7645 \times 10^{-3}$ \\
-- & RBM & $4.9871 \times 10^{-4}$ & $6.9994 \times 10^{-2}$ \\
-- & --- & $4.7901 \times 10^{-5}$ & $6.0408 \times 10^{-3}$ \\
\hline
22 & HRBM & $2.9942 \times 10^{-4}$ & $5.1412 \times 10^{-2}$ \\
-- & --- & $1.2890 \times 10^{-5}$ & $2.1435 \times 10^{-3}$ \\
-- & RBM & $4.3830 \times 10^{-4}$ & $6.9452 \times 10^{-2}$ \\
-- & --- & $1.7918 \times 10^{-5}$ & $2.6969 \times 10^{-3}$ \\
\hline
24 & HRBM & $2.8214 \times 10^{-4}$ & $5.3648 \times 10^{-2}$ \\
-- & --- & $1.4744 \times 10^{-5}$ & $2.5943 \times 10^{-3}$ \\
-- & RBM & $3.2570 \times 10^{-4}$ & $5.8616 \times 10^{-2}$ \\
-- & --- & $1.3413 \times 10^{-5}$ & $2.0967 \times 10^{-3}$ \\
\hline
\end{tabular}
\caption{The mean relative energy error and mean variance of HRBM \& RBM NQS, obtained by averaging over all NQS and QSK random seeds, for different QSK sizes $N=14$ to $N=24$. For each NQS, the mean value is recorded in the first line followed by the standard error in the second line. }\label{main_rel_energy_error}
\end{table}

\begin{table}[!ht]
\centering
\begin{tabular}{c c c c c c}
\hline\hline
 $N$  & NQS &  Mean $S_{vN}$  & Mean $S_2$ & Mean $|S^\text{err}_{vN}|$ & Mean $|S^\text{err}_{2}|$ \\
\hline\hline
14 & Exact & 0.520078 & 0.243722 & -- & -- \\
-- & --- & 0.031519 & 0.020063 & -- & -- \\
-- & HRBM & 0.515985 & 0.242472 & $4.0938 \times 10^{-3}$ & $1.4475 \times 10^{-3}$ \\
-- & --- & 0.031404 & 0.020113 & $8.3657 \times 10^{-4}$ & $4.8295 \times 10^{-4}$ \\
-- & RBM & 0.515263 & 0.241205 & $4.8687 \times 10^{-3}$ & $2.5404 \times 10^{-3}$ \\
-- & --- & 0.031716 & 0.020085 & $7.7701 \times 10^{-4}$ & $4.7317 \times 10^{-4}$ \\
\hline
16 & Exact & 0.593420 & 0.279929 & -- & -- \\
-- & --- & 0.022619 & 0.018099 & -- & -- \\
-- & HRBM & 0.589384 & 0.278968 & $4.0973 \times 10^{-3}$ & $1.3965 \times 10^{-3}$ \\
-- & --- & 0.022729 & 0.018156 & $5.1994 \times 10^{-4}$ & $3.3252 \times 10^{-4}$ \\
-- & RBM & 0.587179 & 0.276833 & $6.2409 \times 10^{-3}$ & $3.0962 \times 10^{-3}$ \\
-- & --- & 0.022567 & 0.018040 & $5.0927 \times 10^{-4}$ & $2.3304 \times 10^{-4}$ \\
\hline
18 & Exact & 0.697194 & 0.342276 & -- & -- \\
-- & --- & 0.016812 & 0.018038 & -- & -- \\
-- & HRBM & 0.691855 & 0.340446 & $5.3549 \times 10^{-3}$ & $2.2193 \times 10^{-3}$ \\
-- & --- & 0.016686 & 0.017588 & $5.0011 \times 10^{-4}$ & $5.9640 \times 10^{-4}$ \\
-- & RBM & 0.691047 & 0.338721 & $6.2518 \times 10^{-3}$ & $3.6799 \times 10^{-3}$ \\
-- & --- & 0.016906 & 0.018100 & $6.6851 \times 10^{-4}$ & $3.8539 \times 10^{-4}$ \\
\hline
20 & Exact & 0.746058 & 0.361794 & -- & -- \\
-- & --- & 0.034468 & 0.028013 & -- & -- \\
-- & HRBM & 0.740603 & 0.359897 & $5.4550 \times 10^{-3}$ & $2.1246 \times 10^{-3}$ \\
-- & --- & 0.034004 & 0.027558 & $7.7750 \times 10^{-4}$ & $6.5151 \times 10^{-4}$ \\
-- & RBM & 0.738240 & 0.357391 & $7.8175 \times 10^{-3}$ & $4.4079 \times 10^{-3}$ \\
-- & --- & 0.034721 & 0.028102 & $1.1843 \times 10^{-3}$ & $7.4506 \times 10^{-4}$ \\
\hline
22 & Exact & 0.879077 & 0.438192 & -- & -- \\
-- & --- & 0.024809 & 0.023930 & -- & -- \\
-- & HRBM & 0.873150 & 0.436089 & $5.9273 \times 10^{-3}$ & $2.4967 \times 10^{-3}$ \\
-- & --- & 0.024534 & 0.023211 & $5.0707 \times 10^{-4}$ & $1.1224 \times 10^{-3}$ \\
-- & RBM & 0.870381 & 0.432984 & $8.6962 \times 10^{-3}$ & $5.2078 \times 10^{-3}$ \\
-- & --- & 0.024750 & 0.023845 & $6.8511 \times 10^{-4}$ & $5.1456 \times 10^{-4}$ \\
\hline
24 & Exact & 0.941250 & 0.469178 & -- & -- \\
-- & --- & 0.036312 & 0.032564 & -- & -- \\
-- & HRBM & 0.933566 & 0.466771 & $7.6845 \times 10^{-3}$ & $2.4120 \times 10^{-3}$ \\
-- & --- & 0.035862 & 0.032399 & $1.0879 \times 10^{-3}$ & $5.8159 \times 10^{-4}$ \\
-- & RBM & 0.933316 & 0.464507 & $7.9337 \times 10^{-3}$ & $4.6705 \times 10^{-3}$ \\
-- & --- & 0.036440 & 0.032625 & $3.6607 \times 10^{-4}$ & $2.1982 \times 10^{-4}$ \\
\hline
\end{tabular}
\caption{This table lists the mean R\'enyi-2 entropy $S_2$, the von Neumann entropy $S_{vN}$ as well as the absolute entropy errors  $|S^\text{err}_{vN}|$, $|S^\text{err}_{2}|$  of HRBM \& RBM NQS, obtained by averaging over all NQS and QSK seeds (corresponding to 12 different QSK disorder realizations) for different QSK sizes $N$. The exact values are the mean over QSK seeds of the exact $S_2$ and $S_{vN}$  from the exact wavefunction obtained by ED.  For each NQS, the mean value is recorded in the first line followed by the standard error  in the second line. For the exact wavefunction, there is no $|S^\text{err}_{2}|$, $|S^\text{err}_{vN}|$ (where the corresponding entries are filled with a `$-$').}\label{main_entropy}
\end{table}

\FloatBarrier
Across all system sizes ($N = 14 \to 24$), the scaling benchmark demonstrates a clear separation between energy optimization, $S_2$ and $S_{vN}$ entanglement entropies, and fine-grained quantum state reconstruction. While both NQS architectures track macroscopic entanglement trends, HRBM systematically suppresses error growth as the Hilbert space dimension expands from $2^{14}$ to $2^{24}$. In particular, the following observations can be made regarding the key metrics.
\bei
\item In terms of the ground state energy optimization ($\varepsilon$ \& $\sigma_E^2$):
\bei 
\item Relative energy error $\varepsilon= \langle |E_{\text{NQS}} - E_{\text{ED}}| / |E_{\text{ED}}| \rangle$ versus $N$:
HRBM maintains systematic superiority across all system sizes. Standard Euclidean RBM suffers from a non-monotonic optimization spike, starting from $3.35 \times 10^{-4}$ at $N=14$ and peaking at $4.99\times 10^{-4}$ at $N=20$. On the other hand, HRBM displays smooth, bounded behavior ($\varepsilon \le 3.0 \times 10^{-4}$). At $N=24$, HRBM reaches $\varepsilon = 2.82 \times 10^{-4}$ compared to RBM's $3.26 \times 10^{-4}$.
\item Energy variance $\sigma_E^2$ versus $N$:  HRBM exhibits lower local energy variance across the entire scaling range. While RBM variance fluctuates non-monotonically (peaking at $\sigma^2 \approx 0.069$ at $N=22$), HRBM scales smoothly from approximately  $0.022$ ($N=14$) to approximately $0.054$ ($N=24$), demonstrating greater physical stability against disorder.

\eni
\item In terms of the macroscopic entanglement entropies ($S_2$ \& $S_{vN}$ versus $N$): Both HRBM and RBM
 display near-perfect overlap with the exact ground state across all system sizes, with HRBM visibly maintaining a closer overlap with the exact state.  Despite this, it must be noted that the $S_2$ and $S_{vN}$ scalar entropies are heavily weighted by the few largest eigenvalues ($\lambda_i > 10^{-4}$), so the macroscopic plots of $S_2$ \& $S_{vN}$ tend to mask underlying NQS architectural differences in representing high-rank correlations.

\item In terms of the absolute entanglement error scaling ($|S_2^{\text{err}}|$ \& $|S_{vN}^{\text{err}}|$ versus $N$):  HRBM achieves lower absolute reconstruction errors. At $N=24$, while the von Neumann entropy errors are comparable for both NQS models with HRBM's being slightly lower at $7.68\times 10^{-3}$ versus RBM's $7.93\times 10^{-3}$,  HRBM reduces the Rényi-2 entropy error ($|S_2 - S_{2,\mathrm{exact}}|$) to $\approx 2.41 \times 10^{-3}$ (almost 2 times lower), compared to $\approx 4.67 \times 10^{-3}$ for standard RBM .
\eni
Additionally, in terms of the individual QSK disorder realizations (see Figs.\ref{scaling_s1111}-\ref{scaling_s9999} in Section \ref{scaling_seeds}), we note the following trends.
\bei
\item  
Each specific $J_{ij}$ coupling matrix (corresponding to a distinct QSK seed) produces a completely different entropy landscape. For instance, looking at various $S_2$ and $S_{vN}$ curves, it can be seen that seed 2222 has a massive spike at $N=18$ and $N=22$, seed 5678 peaks at $N=16$, and seed 7777 spikes at $N=22$.  In every single seed, HRBM and RBM track the exact ground state entropy profile almost identically. 

\item  HRBM consistently achieves lower relative energy errors $\varepsilon$ across virtually all system sizes and seeds. HRBM errors generally remain bounded within $1.0 \times 10^{-4}$ to $4.5 \times 10^{-4}$. On the other hand, Euclidean RBM exhibits higher error baselines and larger variance spikes, reaching relative errors of $5.0 \times 10^{-4}$ to $9.0 \times 10^{-4}$ in challenging realizations such as seed 1234, seed 3333, and seed 9012.

\item  HRBM maintains systematically lower energy variance across all system sizes $N \in [14, 24]$. HRBM variance typically scales smoothly approximately between $0.02$ and $0.05$.  Euclidean RBM variance regularly spikes up to $0.08–0.14$ at intermediate system sizes (e.g., $N=20$ in seed 1234 and seed 3333; $N=22$ in seed 4444 and seed 9012). Lower variance in HRBM is indicative of the fact that the wave function representation stays closer to a true energy eigenstate during VMC optimization.

\item In terms of the absolute entropy errors ($|S - S^{\text{Exact}}|$):
\ben
\item
R\'enyi-2 entropy ($S_2$) error $|S _2- S_2^{\text{Exact}}|$: HRBM maintains tight, stable error boundaries (typically smaller than $0.003$) across $N \in [14, 24]$. Euclidean RBM shows higher error fluctuations, often peaking near $0.008–0.010$ (e.g., seed 1234, seed 3333, seed 4444).

\item von Neumann entropy ($S_{vN}$) error $|S_{vN} - S_{vN}^{\text{Exact}}|$: HRBM errors remain consistently low (around $0.002–0.006$) across most seeds. In isolated configurations with high local disorder variance\footnote{There are a few isolated instances in which HRBM underperforms RBM in terms of various metrics. For instance, at $N=24$, for seed 7777 and seed 9999,  RBM achieves slightly lower relative energy error than HRBM at $N=24$. For seed 2222 at $N=18$, seed 5555 at $N=24$, seed 5678 at $N=20$, seed 7777 at $N=22$, seed 8888 at $N=20, 22$, seed 9999 at $N=18, 20$, HRBM exhibits higher $S_2$ and $S_{vN}$ reconstruction errors than RBM.}, HRBM error bars widen, but overall mean error remains superior or comparable to standard RBM.
\enn
\eni
\subsection{Full entanglement spectrum results} \label{full_entanglement_results}
\begin{figure}[!ht]
\centering
\includegraphics[width=.9\textwidth]{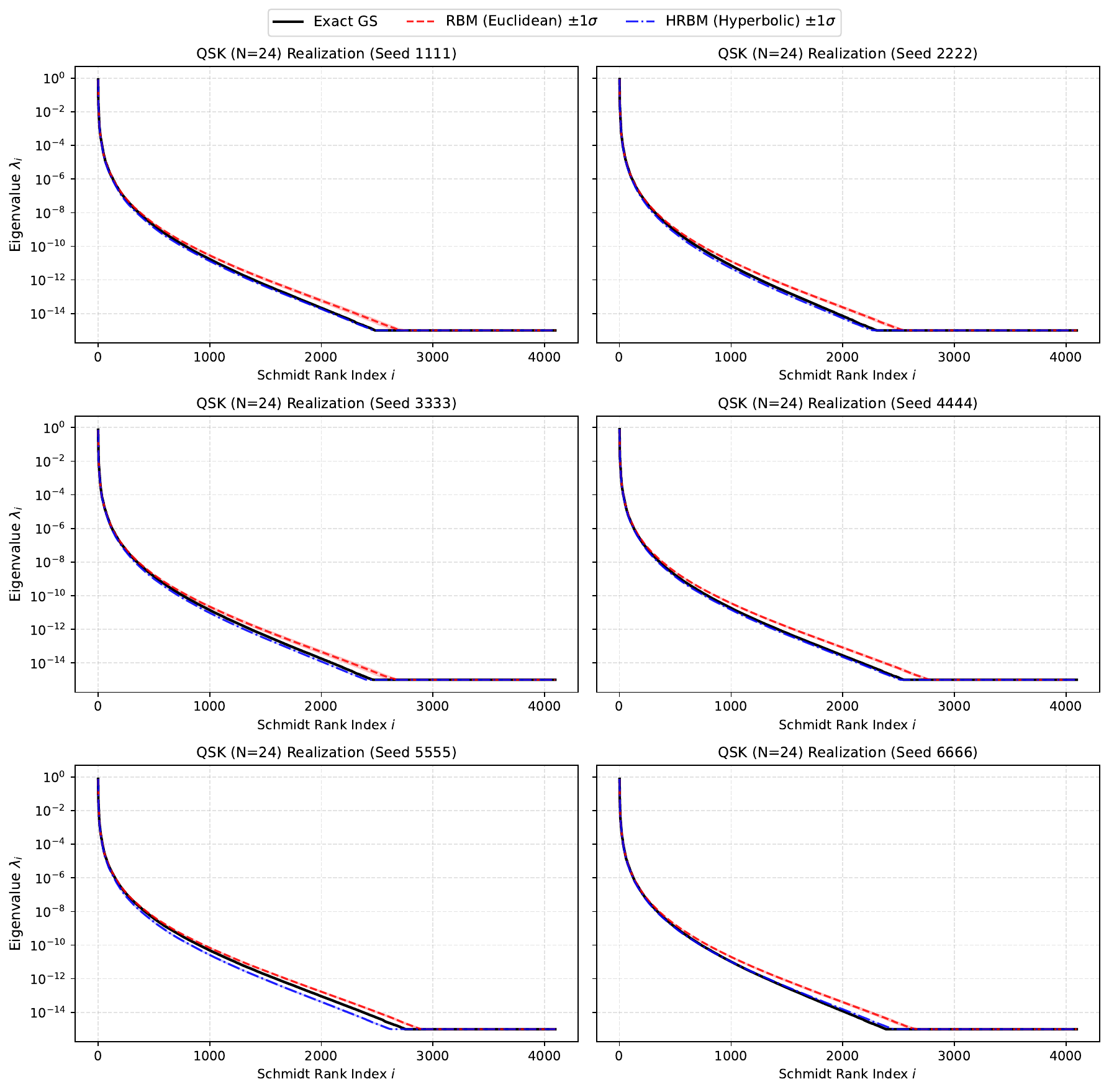}
\caption{Full entanglement spectra of RBM and HRBM NQS in QSK VMC setting with $N=24$ at different QSK model random seeds (1111, 2222, 3333, 4444, 5555, 6666). In each subplot, the RBM and HRBM lines denote the mean value obtained by averaging over different VMC runs involving different NQS random seeds, with shading indicating one standard deviation.}\label{ent_spec_N24_1}
\end{figure}

\begin{figure}[!ht]
\centering
\includegraphics[width=.9\textwidth]{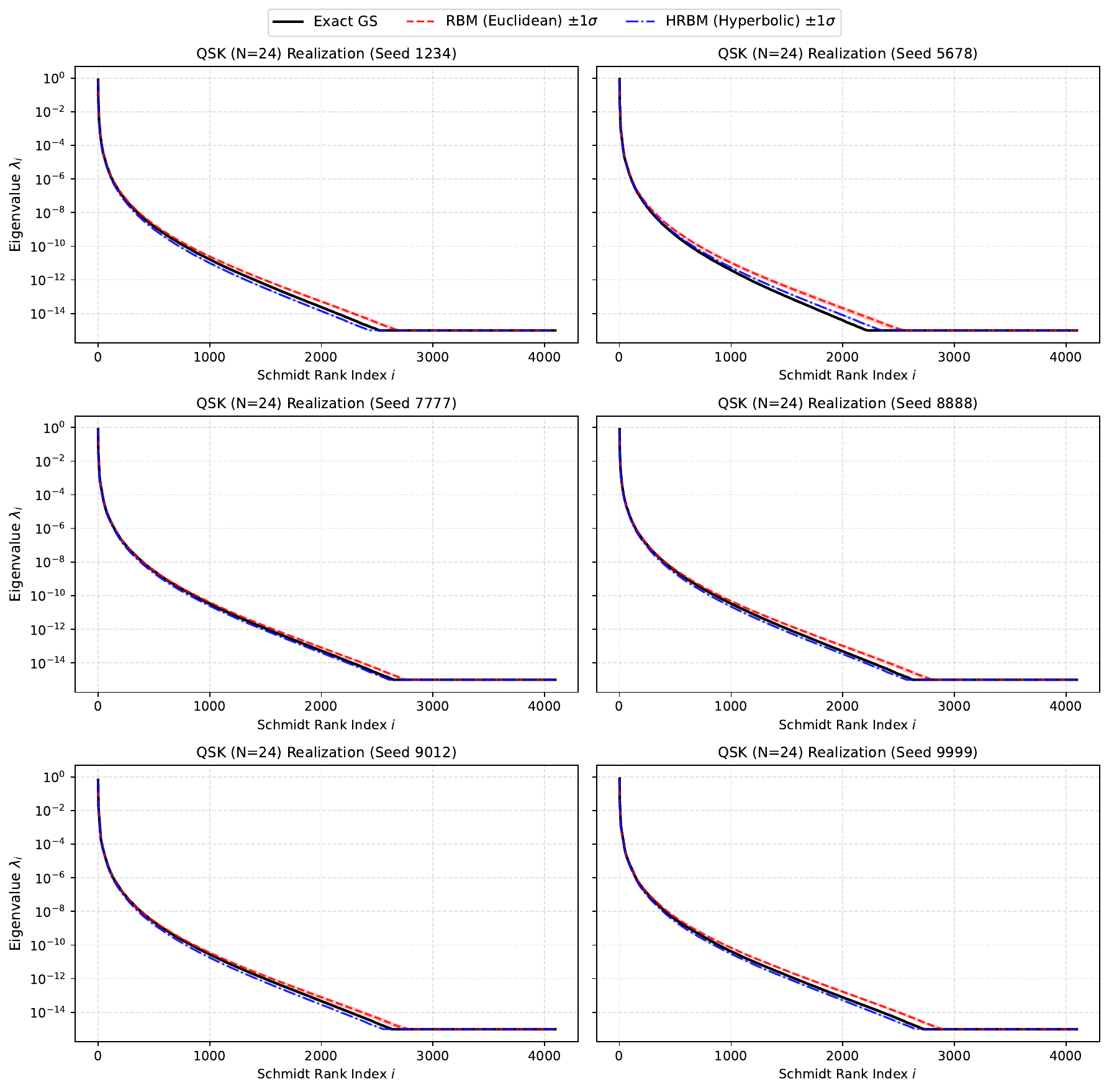}
\caption{Full entanglement spectra of RBM and HRBM NQS in QSK VMC setting with $N=24$ at different QSK model random seeds (1234, 5678, 7777, 8888, 9012, 9999). In each subplot, the RBM and HRBM lines denote the mean value obtained by averaging over different VMC runs involving different NQS random seeds, with shading indicating one standard deviation.}\label{ent_spec_N24_2}
\end{figure}

 In the previous section, we analyzed scalar metrics like energy and integrated entropies, but these metrics do not provide the full picture as macroscopic scalar entropies like $S_2$ and  $S_{vN}$ mask critical model deficiencies given that they are overwhelmingly dominated by the top few eigenvalues of the reduced density matrix $\rho$ (where $\lambda_i > 10^{-4}$), so a model can achieve respectable relative energy errors while failing to capture the true quantum state structure.
On the other hand, the full entanglement spectrum ($\lambda_i$) - in particular the spectral tail decay - is much more informative in terms of identifying the true expressivity of an NQS model.  As such, in this section, we discuss the full entanglement spectrum $\lambda_i$ results of the VMC experiments done in order to comprehensively compare the performance of HRBM versus RBM NQS. 
\\\\
The plots showing the full entanglement spectra for $N=24$ at 12 different QSK disorder realizations are included in Fig.\ref{ent_spec_N24_1} and Fig.\ref{ent_spec_N24_2} while those for $N=14$ to $N=22$ are included in Fig.\ref{ent_spec_N14_1} - Fig.\ref{ent_spec_N22_2} in the Appendix section \ref{ent_plots}. 
\FloatBarrier
\FloatBarrier

Based on Figs.\ref{ent_spec_N24_1}-\ref{ent_spec_N22_2}, the following observations can be made:
\bei
\item \textit{With respect to the top eigenvalues (where $10^{-6}<\lambda_i < 10^0 $)}: Both RBM and HRBM match the exact ground state curve (black) almost perfectly across all system sizes and disorder seeds. Because $S_2$ and $S_{vN}$ are heavily weighted by these largest eigenvalues ($\lambda_i^2$ for $S_2$ and $-\lambda_i \log \lambda_i$ for $S_{vN}$), both models appear to coincide with the exact result on macro-level plots.
\item \textit{With respect to the sub-dominant tail (where eigenvalues $\lambda_i < 10^{-8}$)}, the true representation bottleneck of Euclidean space is revealed.
\bei
\item RBM (the red dashed line) systematically overestimates the tail eigenvalues across virtually every QSK seed and size. The RBM line visibly drifts above the exact curve, creating a ``fat tail" that accumulates spurious weight in high-index Schmidt modes.
Because flat Euclidean space expands polynomially, standard RBM lacks the geometric capacity to enforce sharp eigenmode truncation during VMC optimization, which probably leads to the over-parameterization of sub-dominant modes.
\item  HRBM (the blue dash-dotted line) faithfully tracks the steep decay of the exact ground state curve (in black) down to the numerical noise floor ($\sim 10^{-15}$) for most of the QSK disorder realizations, showing almost no upward drift, thanks to hyperbolic geometry's exponential volume matching the hierarchical decay of the entanglement spectrum. For some of the QSK seeds, the HRBM curve actually underestimates the exact curve. This is probably because the exponential volume expansion in hyperbolic space enforces a strict hierarchical structure. In disorder realizations where Schmidt modes decay smoothly rather than hierarchically, HRBM slightly over-compresses higher-order Schmidt modes, leading to an under-estimation in the tail ($\lambda_i^{\text{HRBM}} < \lambda_i^{\text{Exact}}$). On the other hand, flat Euclidean geometry cannot scale volume exponentially, so to represent non-local spin-glass correlations, RBM spreads amplitude across spurious product states, leading to a systematic over-estimation of lower Schmidt eigenvalues ($\lambda_i^{\text{RBM}} > \lambda_i^{\text{Exact}}$).
\eni
\item \textit{With respect to the scaling property with system size $N$}: As $N$ increases from 14 to 24 (where the number of eigenmodes grows to beyond $4000$), RBM's spectral deviation in the tail becomes progressively wider, while HRBM stays tightly bound to the exact curve. This shows that HRBM ansatz genuinely captures the underlying quantum state topology.
\eni
 Given the results of this proof-of-concept work, it would be interesting to scale the QSK system up beyond $N=24$ and carry out the same study done in this work. However, at $N=26$, exact diagonalization becomes much harder than at $N=24$, given the doubling of the Hilbert space dimension, so for $N\geq 26$, alternative computational methods are required to provide a reliable benchmark to test the performances of the NQS. Also, with our current computing resources, while it is not feasible to carry out the computations for the $N=26$ case, it is natural to wonder whether, at $N=26$, HRBM would continue to outperform RBM, given the trends shown in Fig.\ref{main_scaling_res}, where RBM displays a non-monotonic optimization trajectory with an error spike at $N=20$ whereas HRBM displays a monotonic trajectory. Could this mean that at $N=26$, RBM would outperform HRBM in terms of relative energy error, energy variance and reconstruction errors? To answer this question (speculatively), we rely on the following observations. First, the RBM non-monotonicity is characteristic of optimization instability in flat geometry: As $N$ scales, Euclidean landscapes develop complex local minima traps where optimization success varies unpredictably depending on system size and disorder realization. On the other hand, the smooth scaling and strict monotonicity of HRBM indicate that hyperbolic geometry provides a consistent, predictable representation capacity that does not suffer from geometric optimization bottlenecks as Hilbert space expands. Second, and more crucially, even when RBM approaches HRBM in the scalar metrics like energy and $S_2$ or $S_{vN}$ entropies at $N=24$, it still fails to capture non-local quantum correlations, maintaining an unphysical `fat tail' (where $\lambda_i< 10^{-8}$) in its entanglement spectrum for all QSK realizations at all sizes, in direct constrast to HRBM's ability to reconstruct the entire spectrum down to $\lambda_i \sim 10^{-15}$. This means that even at $N=26$ or beyond, HRBM would likely emerge as the better NQS variant to represent the QSK volume-law ground state wavefunction, given its capacity to faithfully match the hierarchical entanglement spectrum decay. 
\section{Concluding remarks} \label{concl}
In this work, we have introduced the first type of non-autoregressive non-Euclidean NQS ansatz in the form of the hyperbolic RBM (HRBM) NQS, which is shown to be capable of robustly outperforming the Euclidean RBM NQS in the volume-law QSK system representing quantum spin glasses. In a scaling study where  both architectures were evaluated under identical polynomial parameter budgets ($O(\text{poly}(N))$), we tracked the performances of the RBM and HRBM NQS across an expansion of $2^{10} = 512$ folds in the Hilbert space dimension for QSK models whose size $N$ ranges from $N=14$ to $N=24$. These small sizes were chosen because exact diagonalization is feasible computationally and provides an exact benchmark against which to compare the performances of the RBM and HRBM NQS.
\\\\
In terms of both the reachable ground state energy and entanglement entropies (including the $S_2$ and von-Neumann $S_{vN}$), HRBM outperforms RBM across all QSK size $N$ as shown in Fig.\ref{main_scaling_res}.
While standard Euclidean RBM displays a non-monotonic optimization trajectory as $N$ scales, HRBM achieves asymptotic convergence across both energy and entropy metrics.
 Even more interestingly, HRBM also demonstrates its ability to faithfully capture the full entanglement spectrum of the QSK model across 15 orders of magnitudes across all system sizes, while RBM visibly overestimates the spectrum for sub-dominant modes  (see Figs.\ref{ent_spec_N24_1}-\ref{ent_spec_N22_2}). This is very likely due to the fundamental differences in the geometry underlying the two types of NQS. Since Euclidean latent space grows only polynomially, standard Euclidean RBM lacks the geometric capacity to enforce sharp, high-order entanglement mode decay.  On the other hand, the fact that quantum entanglement spectra decay exponentially ($\lambda_i \sim e^{-\alpha i}$ for some proportionality constant $\alpha$) means that the exponential volume expansion of hyperbolic space naturally mirrors this hierarchical entanglement eigenvalue decomposition, allowing HRBM to represent thousands of sub-dominant modes without parameter saturation or spectral leakage. 
 This is arguably the strongest supporting evidence for our hypothesis that hyperbolic NQS (including the HRBM considered in this work as well as other types of yet-to-be-constructed hyperbolic NQS ans\"atze) might be a genuinely better match than Euclidean NQS for volume-law quantum systems thanks to their superior expressivity.
Furthermore, we note that an interesting byproduct of the results of this work is the numerical proof showing that in order to maintain a relative energy error of less than $10^{-3}$, only polynomial scaling of the RBM/HRBM NQS is required\footnote{From $N=14$ to $N=24$, the number of parameters of RBM/HRBM NQS only increases by a factor of less than 3, from 224 at $N=14$ to 624 at $N=24$ (see Table \ref{vmc_params}), while the Hilbert space dimension increases by 512 times.} when the Hilbert space dimension of the volume-law QSK system expands exponentially from $2^{14}$ to $2^{24}$ states.
\\\\
Given the results of this proof-of-concept work that successfully demonstrate the potential of non-autoregressive hyperbolic RBM NQS to achieve high physical fidelity for volume-law states without requiring exponential parameter expansion, it would be instructive to construct new types of hyperbolic NQS such as hyperbolic transformer NQS or hyperbolic graph neural network NQS and study these new constructions in the context of volume-law quantum systems. Another promising future direction might involve combining the HRBM NQS constructed in this work with a Slater determinant to tackle the highly challenging volume-law fermionic SYK model. More concretely, in fermionic systems, the Slater determinant enforces the required anti-symmetry under particle exchange, while the HRBM network acts as an efficient non-local correlation backflow factor. Ideally, pairing the two should leverage the geometric efficiency of hyperbolic space to capture strong, dense electron correlations that standard Euclidean neural-network backflow factors struggle to represent without exponential parameters. 
We hope to return to these in future works. 
\appendix
\section{Appendix}
\subsection{Hyperbolic Lorentzian mathematical operations}  \label{lor_math}
The $n$-dimensional Lorentz hyperboloid $\mathbb H^n$ with constant negative curvature $-k$  is given as
\beq
\mathbb H^n \equiv \left\{\mathbf x\in \mathbb R^{n+1}: \langle \mathbf x,\mathbf x\rangle_{\mathcal L} = -k, \,x_0>0\rr\}
\eeq
where $\langle \mathbf x, \mathbf y \rangle_{\mathcal L}$, the Lorentzian scalar product for $(n+1)$-dimensional vectors $\mathbf{x, y} \in \mathbb R^{n+1}$, is defined as
\beq
\langle \mathbf x, \mathbf y \rangle_{\mathcal L} \equiv -x_0y_0 + \sum_{i=1}^n x_iy_i\,.
\eeq
Note that in the Lorentz model of hyperbolic space, the origin is the point $\mathbf 0_{\mathcal L} = (1, 0, 0,\ldots, 0)$.
Throughout this work, we fix $k=1$. All equations that follow has $k=1$. 
\\\\
The distance function between two points $\mathbf x, \mathbf y \in \mathbb H^n$ is 
\beq
d_{\mathbb H}(\mathbf x, \mathbf y) = \text{arcosh}\left(-\langle  \mathbf x, \mathbf y\rangle_{\mathcal L}\rr)\,,
\eeq
while the Lorentzian norm of a vector $\mathbf v$ is 
\beq
||\mathbf v||_{\mathcal L} = \sqrt{\langle  \mathbf v, \mathbf v\rangle_{\mathcal L}}\,.
\eeq
The tangent space at $\mathbf x$ is the $n$-dimensional Euclidean vector space approximating $\mathbb H^n$ around $\mathbf x$:
\beq
\mathcal T_x\mathbb H^n \equiv \lf\{ \mathbf x\in \mathbb R^{n+1}: \langle \mathbf v, \mathbf x\rangle_{\mathcal L} = 0\rr\}
\eeq
Mappings between the Lorentz hyperboloid and its tangent space are done using the exponential $\exp_{\mathbf x}(\mathbf v)$ and logarithmic $\log_{\mathbf{x}}(\mathbf y)$ maps.
\beq
\mathcal T_{\mathbf x} \mathbb H^n \rightarrow \mathbb H^n: &&\exp_{\mathbf x}(\mathbf v) = \cosh\lf(||\mathbf v||_{\mathcal L}\rr) \mathbf x + \sinh\lf(||\mathbf v||_{\mathcal L}\rr)\frac{\mathbf v}{||\mathbf v||_{\mathcal L}} \label{eq-l-expmap}\\
\mathbb H^n \rightarrow \mathcal T_{\mathbf x} \mathbb H^n: && \log_{\mathbf{x}}(\mathbf y) =d_{\mathbb H}(\mathbf x, \mathbf y) \frac{\mathbf y + \langle \mathbf x, \mathbf y\rangle_{\mathcal L}\,\mathbf x}{\left||\mathbf y + \langle \mathbf x, \mathbf y\rangle_{\mathcal L}\,\mathbf x\right||_{\mathcal L}} \label{eq-l-logmap}
\eeq
The parallel transport operation that maps a point $\mathbf{z} \in \mathcal T_{\mathbf x} \mathbb H^n$ to a point in $\mathcal T_{\mathbf y}\mathbb H^n$ is defined as
\beq
P_{\mathbf{x}\ra \mathbf{y}}(\mathbf{z}) = \mathbf z+\frac{\langle \mathbf y, \mathbf z\rangle_{\mathcal L}}{1-\langle(\mathbf x, \mathbf y)\rangle_{\mathcal L}}(\mathbf x+ \mathbf y)
\eeq
When $\mathbf x= \mathbf 0_{\mathcal L}$, 
\beq
P_{\mathbf 0_{\mathcal L} \ra \mathbf{y}}(\mathbf z) = \mathbf z+\frac{\langle \mathbf y, \mathbf z\rangle_{\mathcal L}}{1-\langle(\mathbf 0_{\mathcal L}, \mathbf y)\rangle_{\mathcal L}}(\mathbf 0_{\mathcal L}+ \mathbf y) \label{ptrans0}
\eeq
The various Lorentz mathematical operations are defined in terms of the exponential/logarithm mappings at $\mathbf x= \mathbf 0_{\mathcal L}$ in (\ref{eq-l-expmap}), (\ref{eq-l-logmap}) and parallel transport operations given in (\ref{ptrans0})  as follows.
\bei
\item For $\mathbf x, \mathbf y \in \mathbb H^n$, the Lorentz addition $ \oplus_{\mathcal L}$ is defined as:
\beq
\text{Lorentz addition}: \qquad \mathbf x \oplus_{\mathcal L} \mathbf y = \exp_{\mathbf x}\lf[P_{\mathbf 0_{\mathcal L} \ra \mathbf{x}}(\log_{\mathbf 0_{\mathcal L}}(\mathbf y))\rr] \label{lor-add}
\eeq
\item The scalar multiplication $\odot_{\mathcal L}$ between a point $\mathbf x\in \mathbb H^n$ and $r\in \mathbb R$ is 
\beq
\text{Lorentz scalar multiplication}: \qquad r \odot_{\mathcal L} \mathbf x =\exp_{\mathbf 0_{\mathcal L}}\lf[ r \log_{\mathbf 0_{\mathcal L}}(\mathbf x)\rr]
\eeq
\item The matrix multiplication $\otimes_{\mathcal L}$ between a point $\mathbf x\in \mathbb H^n$ and a matrix  $M\in \mathbb R^n\times \mathbb R^n$ is 
\beq
M \otimes_{\mathcal L} \mathbf x =\exp_{\mathbf 0_{\mathcal L}}\lf[ M \log_{\mathbf 0_{\mathcal L}}(\mathbf x)\rr] \label{lor-mult}
\eeq
\item The nonlinear activation $f^{\otimes_{\mathcal L}}$(x) where $x\in \mb H^n$ is
\beq
f^{\otimes_{\mathcal L}} = \exp_{\mathbf 0_{\mathcal L}}\lf(f(\log_{\mathbf 0_{\mathcal L}}(x))\rr)
\eeq
\eni
\FloatBarrier
\clearpage
\subsection{QSK models: Exact energy, $S_2$ and $S_{vN}$ entropies} \label{exact_E_S}
\begin{table}[!ht]
\centering
\begin{tabular}{c c c c c }
\hline\hline
 N  & QSK seed &   Exact Energy  &  Exact $S_2$ &  Exact $S_{vN}$ \\
\hline\hline
14 & 1111 & -14.646850 & 0.115322 & 0.304463 \\
14 & 1234 & -14.764654 & 0.238397 & 0.523197 \\
14 & 2222 & -14.955097 & 0.244716 & 0.550411 \\
14 & 3333 & -14.864592 & 0.228835 & 0.510754 \\
14 & 4444 & -14.762866 & 0.241584 & 0.527381 \\
14 & 5555 & -14.662846 & 0.160396 & 0.390808 \\
14 & 5678 & -14.784242 & 0.249586 & 0.541275 \\
14 & 6666 & -14.884761 & 0.300660 & 0.612231 \\
14 & 7777 & -14.859196 & 0.268521 & 0.489171 \\
14 & 8888 & -14.886363 & 0.251490 & 0.575272 \\
14 & 9012 & -14.777944 & 0.223110 & 0.469309 \\
14 & 9999 & -15.419419 & 0.402053 & 0.746667 \\
\hline
16 & 1111 & -16.835985 & 0.286713 & 0.622081 \\
16 & 1234 & -16.992467 & 0.245180 & 0.553498 \\
16 & 2222 & -16.837838 & 0.245873 & 0.558379 \\
16 & 3333 & -16.920466 & 0.248834 & 0.564964 \\
16 & 4444 & -16.946620 & 0.225920 & 0.529474 \\
16 & 5555 & -16.921646 & 0.246530 & 0.568401 \\
16 & 5678 & -17.101175 & 0.406380 & 0.703491 \\
16 & 6666 & -17.079820 & 0.257172 & 0.544865 \\
16 & 7777 & -16.984548 & 0.249126 & 0.540182 \\
16 & 8888 & -17.112805 & 0.323418 & 0.652029 \\
16 & 9012 & -16.823153 & 0.227077 & 0.513519 \\
16 & 9999 & -17.128633 & 0.396925 & 0.770161 \\
\hline
18 & 1111 & -19.029824 & 0.320822 & 0.692728 \\
18 & 1234 & -18.989854 & 0.254455 & 0.593687 \\
18 & 2222 & -19.021023 & 0.374981 & 0.721321 \\
18 & 3333 & -19.182370 & 0.355702 & 0.680803 \\
18 & 4444 & -19.142463 & 0.361591 & 0.718623 \\
18 & 5555 & -19.091613 & 0.326545 & 0.702677 \\
18 & 5678 & -18.995777 & 0.283652 & 0.655164 \\
18 & 6666 & -19.192267 & 0.324922 & 0.665972 \\
18 & 7777 & -19.246342 & 0.395699 & 0.766011 \\
18 & 8888 & -19.179125 & 0.324541 & 0.689151 \\
18 & 9012 & -18.976817 & 0.289690 & 0.656387 \\
18 & 9999 & -19.729642 & 0.494707 & 0.823804 \\
\hline\hline
\end{tabular}
\caption{Exact ground state energy (obtained by exact diagonalization) and exact $S_2$, $S_{vN}$ entropies of the 12 different QSK disorder realizations for QSK system size $N=14, 16, 18$.}\label{exact_metrics_p1}
\end{table}
\begin{table}[!ht]
\centering
\begin{tabular}{c c c c c }
\hline\hline
 N  & QSK seed &   Exact Energy  &  Exact $S_2$ &  Exact $S_{vN}$ \\
\hline\hline
20 & 1111 & -21.232203 & 0.329475 & 0.696017 \\
20 & 1234 & -21.166380 & 0.319844 & 0.714197 \\
20 & 2222 & -21.004620 & 0.251775 & 0.600228 \\
20 & 3333 & -21.086525 & 0.290121 & 0.659335 \\
20 & 4444 & -21.069415 & 0.282516 & 0.652132 \\
20 & 5555 & -21.584556 & 0.540601 & 0.919006 \\
20 & 5678 & -21.245930 & 0.317694 & 0.672997 \\
20 & 6666 & -21.450500 & 0.487732 & 0.909258 \\
20 & 7777 & -21.319349 & 0.382609 & 0.785670 \\
20 & 8888 & -21.088029 & 0.278855 & 0.636075 \\
20 & 9012 & -21.316428 & 0.356613 & 0.762007 \\
20 & 9999 & -21.842648 & 0.503695 & 0.945772 \\
\hline
22 & 1111 & -23.250564 & 0.396493 & 0.863613 \\
22 & 1234 & -23.306757 & 0.340331 & 0.750125 \\
22 & 2222 & -23.351765 & 0.417939 & 0.860267 \\
22 & 3333 & -23.385946 & 0.403358 & 0.835140 \\
22 & 4444 & -23.524750 & 0.542537 & 1.042305 \\
22 & 5555 & -23.433975 & 0.362869 & 0.783808 \\
22 & 5678 & -23.227276 & 0.389616 & 0.840151 \\
22 & 6666 & -23.375017 & 0.374663 & 0.831556 \\
22 & 7777 & -23.689611 & 0.604698 & 0.991434 \\
22 & 8888 & -23.726312 & 0.546313 & 0.978250 \\
22 & 9012 & -23.425041 & 0.427985 & 0.881539 \\
22 & 9999 & -23.806345 & 0.451503 & 0.890738 \\
\hline
24 & 1111 & -25.457370 & 0.374051 & 0.848803 \\
24 & 1234 & -25.424855 & 0.457565 & 0.929380 \\
24 & 2222 & -25.161303 & 0.295176 & 0.711131 \\
24 & 3333 & -25.588078 & 0.558313 & 0.986689 \\
24 & 4444 & -25.582844 & 0.473011 & 0.955019 \\
24 & 5555 & -25.641346 & 0.511163 & 1.043722 \\
24 & 5678 & -25.342593 & 0.305626 & 0.710423 \\
24 & 6666 & -25.910862 & 0.539969 & 1.021905 \\
24 & 7777 & -25.625563 & 0.447296 & 0.959343 \\
24 & 8888 & -25.707300 & 0.479705 & 0.995557 \\
24 & 9012 & -26.044812 & 0.710787 & 1.117530 \\
24 & 9999 & -25.562186 & 0.477472 & 1.015500 \\
\hline\hline
\end{tabular}
\caption{Exact ground state energy (obtained by exact diagonalization) and exact $S_2$, $S_{vN}$ entropies of the 12 different QSK disorder realizations for QSK system size $N=20, 22, 24$.}\label{exact_metrics_p2}
\end{table}
\clearpage
\subsection{Scaling results at different QSK realizations} \label{scaling_seeds}

\begin{figure}[!ht]
\centering
\includegraphics[width=.9\textwidth]{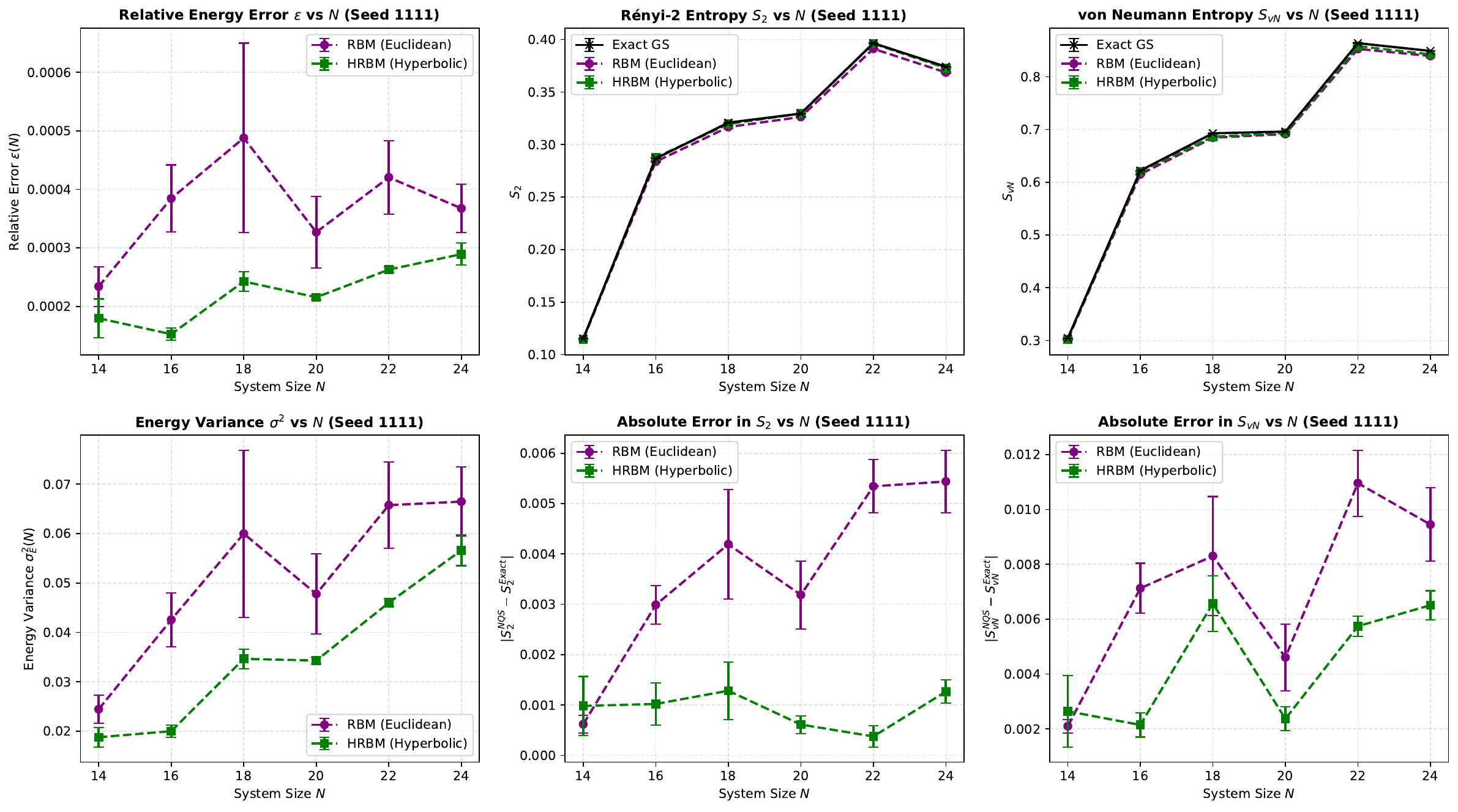}
\caption{Scaling benchmark (QSK seed=1111) for RBM and HRBM in the QSK setting as the system size $N$ increases from $N=14$ to $N=24$. The error bars are obtained by averaging over different NQS random seeds.}\label{scaling_s1111}
\end{figure}
\FloatBarrier

\begin{figure}[!ht]
\centering
\includegraphics[width=.9\textwidth]{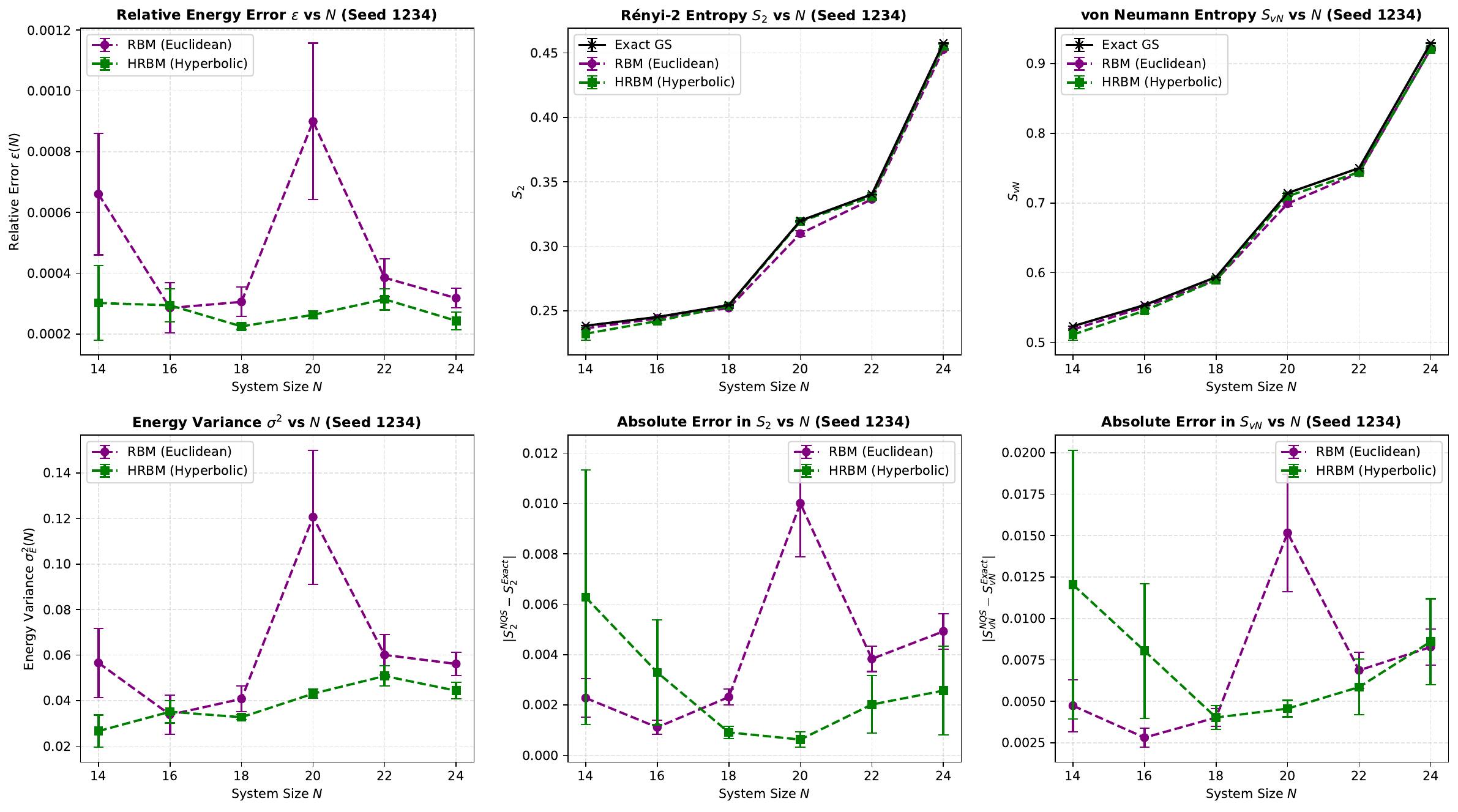}
\caption{Scaling benchmark (QSK seed=1234) for RBM and HRBM in the QSK setting as the system size $N$ increases from $N=14$ to $N=24$. The error bars are obtained by averaging over different NQS random seeds.}\label{scaling_s1234}
\end{figure}
\FloatBarrier

\begin{figure}[!ht]
\centering
\includegraphics[width=.9\textwidth]{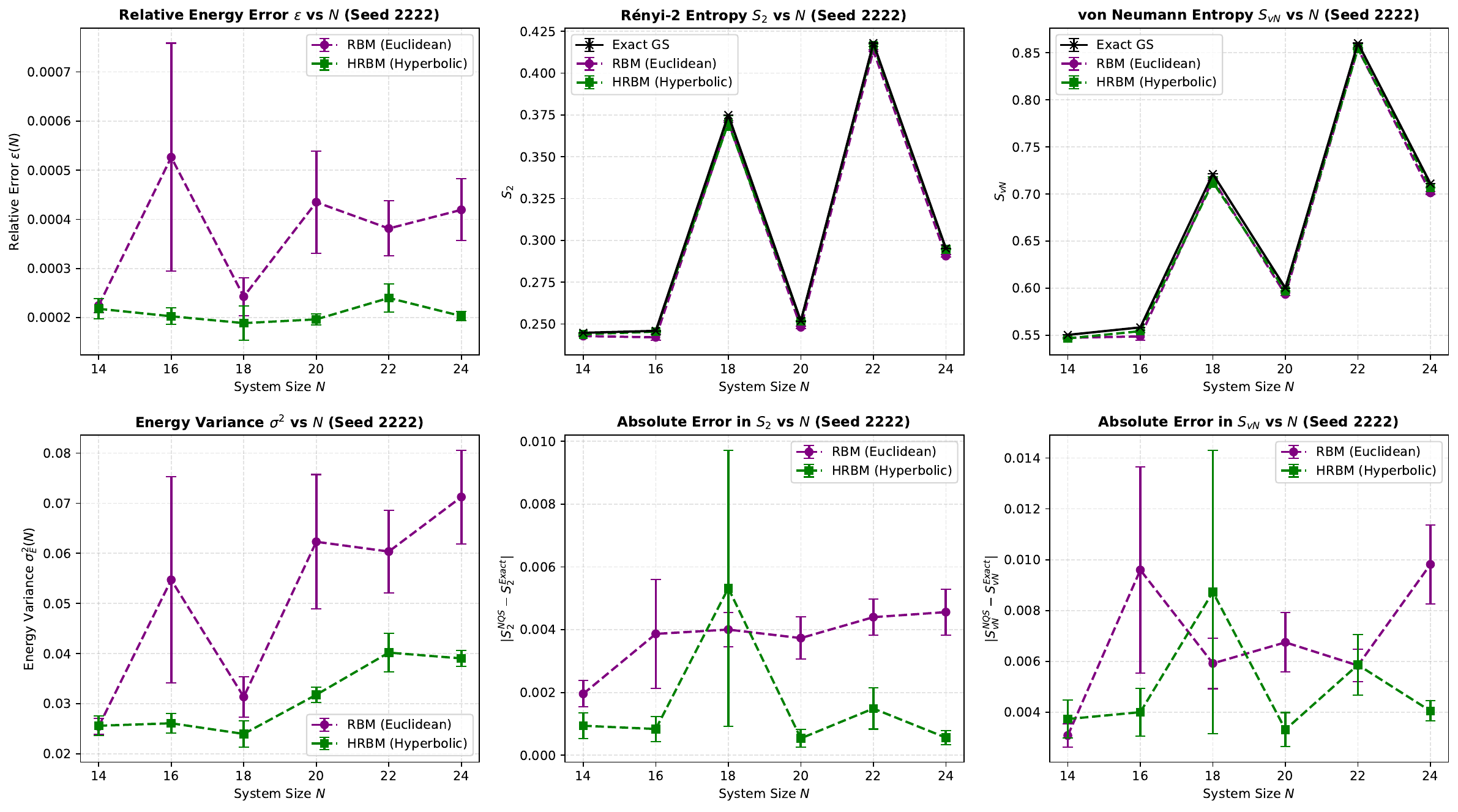}
\caption{Scaling benchmark (QSK seed=2222) for RBM and HRBM in the QSK setting as the system size $N$ increases from $N=14$ to $N=24$. The error bars are obtained by averaging over different NQS random seeds.}\label{scaling_s2222}
\end{figure}
\FloatBarrier

\begin{figure}[!ht]
\centering
\includegraphics[width=.9\textwidth]{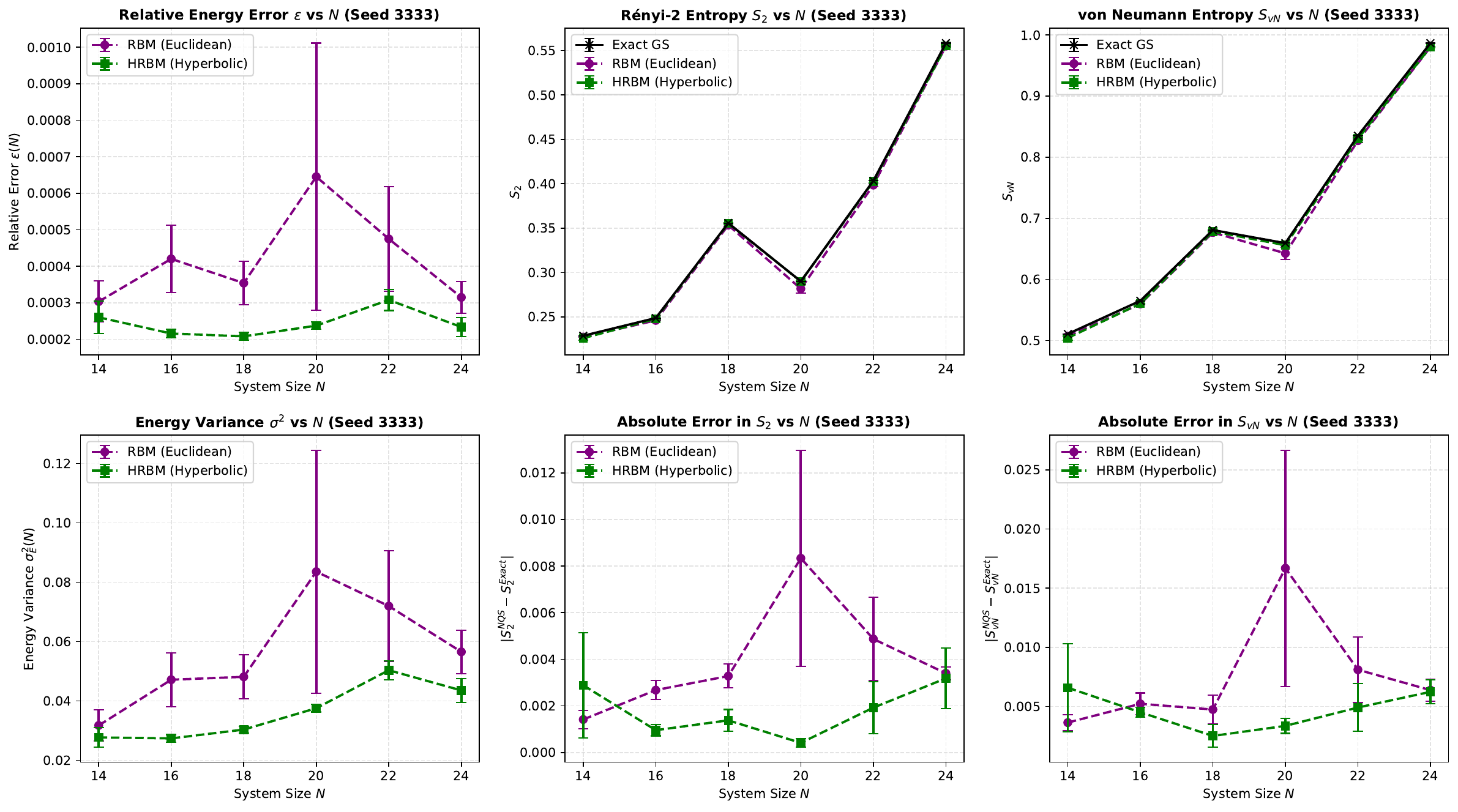}
\caption{Scaling benchmark (QSK seed=3333) for RBM and HRBM in the QSK setting as the system size $N$ increases from $N=14$ to $N=24$. The error bars are obtained by averaging over different NQS random seeds.}\label{scaling_s3333}
\end{figure}
\FloatBarrier

\begin{figure}[!ht]
\centering
\includegraphics[width=.9\textwidth]{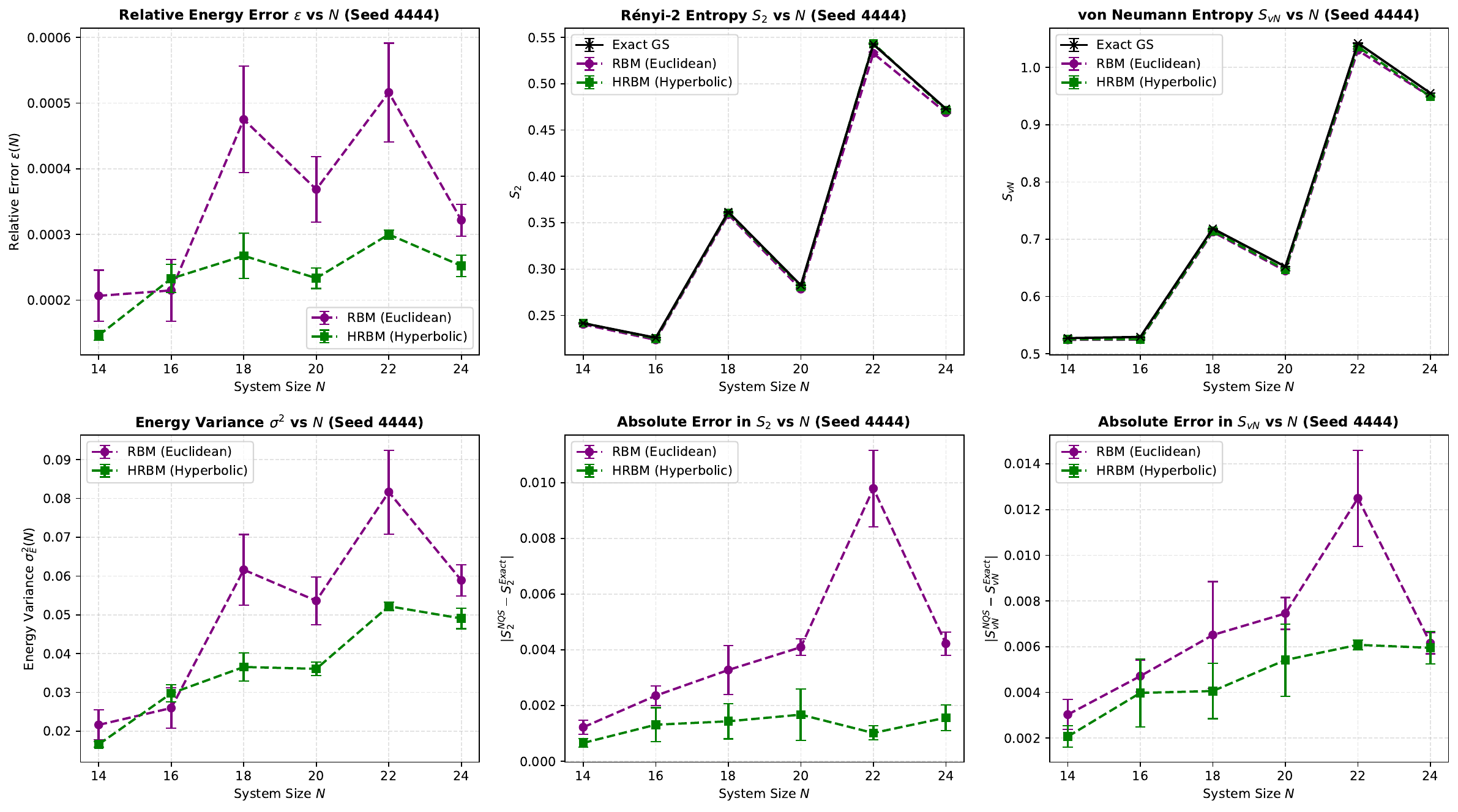}
\caption{Scaling benchmark (QSK seed=4444) for RBM and HRBM in the QSK setting as the system size $N$ increases from $N=14$ to $N=24$. The error bars are obtained by averaging over different NQS random seeds.}\label{scaling_s4444}
\end{figure}
\FloatBarrier

\begin{figure}[!ht]
\centering
\includegraphics[width=.9\textwidth]{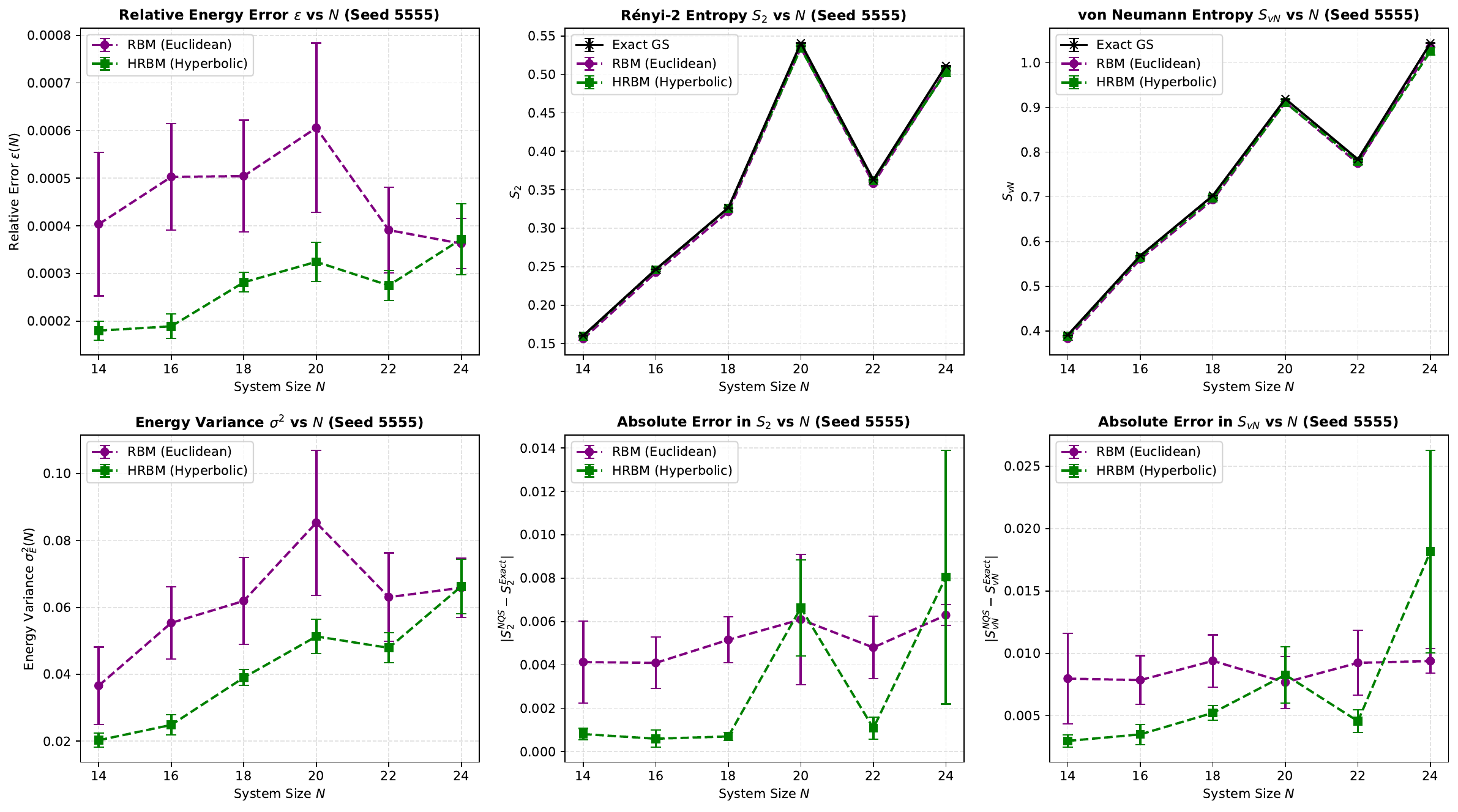}
\caption{Scaling benchmark (QSK seed=5555) for RBM and HRBM in the QSK setting as the system size $N$ increases from $N=14$ to $N=24$. The error bars are obtained by averaging over different NQS random seeds.}\label{scaling_s5555}
\end{figure}
\FloatBarrier
\begin{figure}[!ht]
\centering
\includegraphics[width=.9\textwidth]{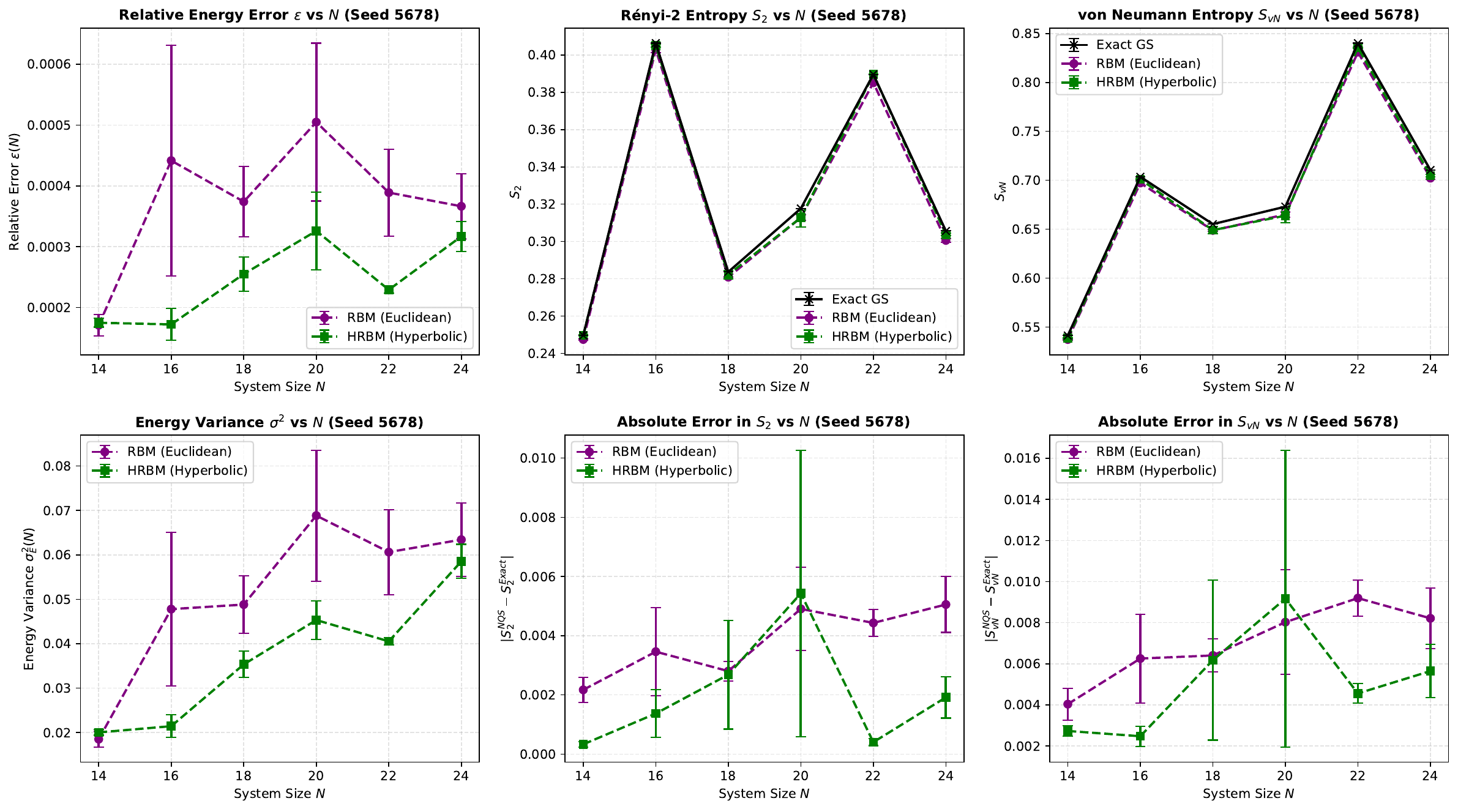}
\caption{Scaling benchmark (QSK seed=5678) for RBM and HRBM in the QSK setting as the system size $N$ increases from $N=14$ to $N=24$. The error bars are obtained by averaging over different NQS random seeds.}\label{scaling_s5678}
\end{figure}
\FloatBarrier

\begin{figure}[!ht]
\centering
\includegraphics[width=.9\textwidth]{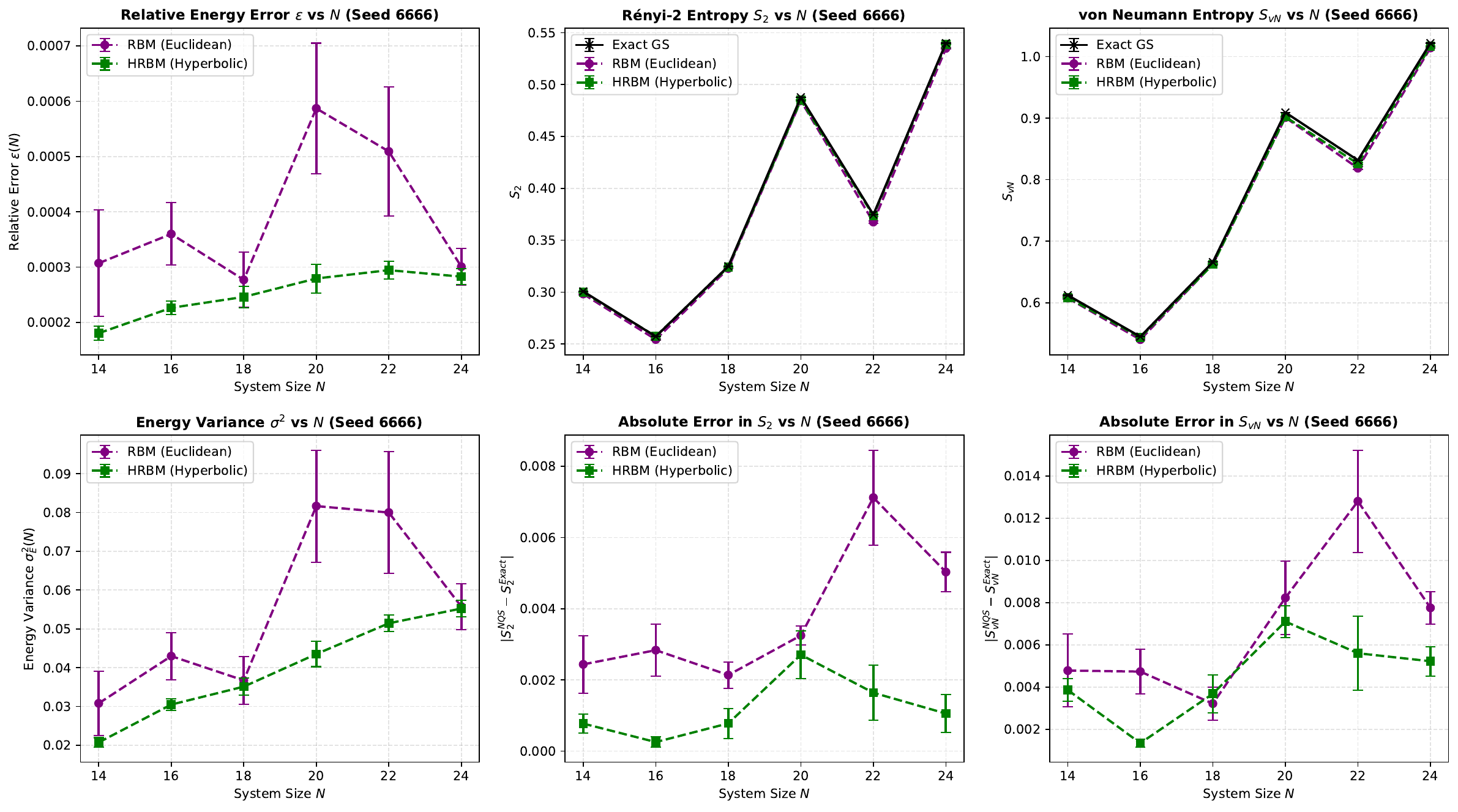}
\caption{Scaling benchmark (QSK seed=6666) for RBM and HRBM in the QSK setting as the system size $N$ increases from $N=14$ to $N=24$. The error bars are obtained by averaging over different NQS random seeds.}\label{scaling_s6666}
\end{figure}
\FloatBarrier
\begin{figure}[!ht]
\centering
\includegraphics[width=.9\textwidth]{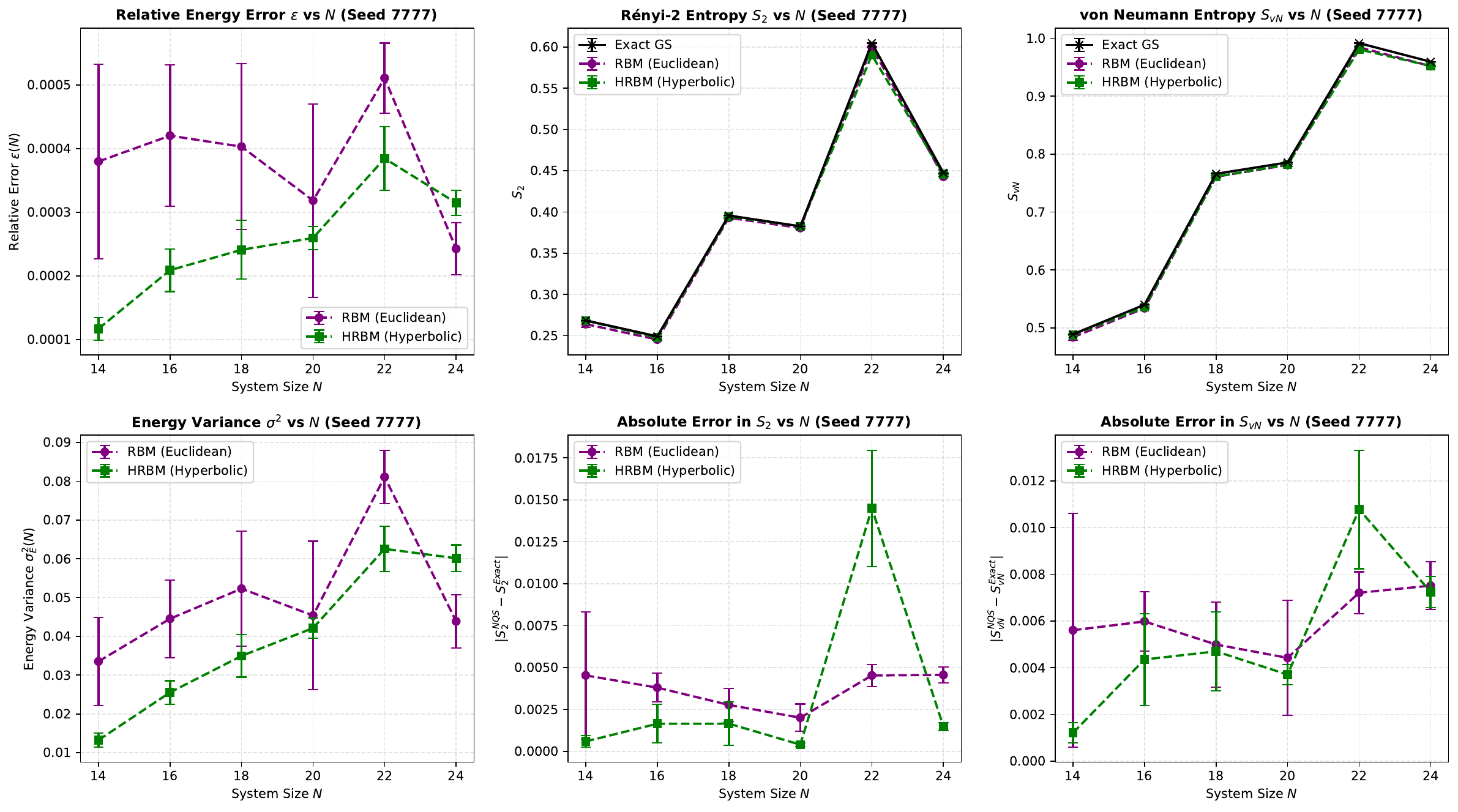}
\caption{Scaling benchmark (QSK seed=7777) for RBM and HRBM in the QSK setting as the system size $N$ increases from $N=14$ to $N=24$. The error bars are obtained by averaging over different NQS random seeds.}\label{scaling_s7777}
\end{figure}
\FloatBarrier

\begin{figure}[!ht]
\centering
\includegraphics[width=.9\textwidth]{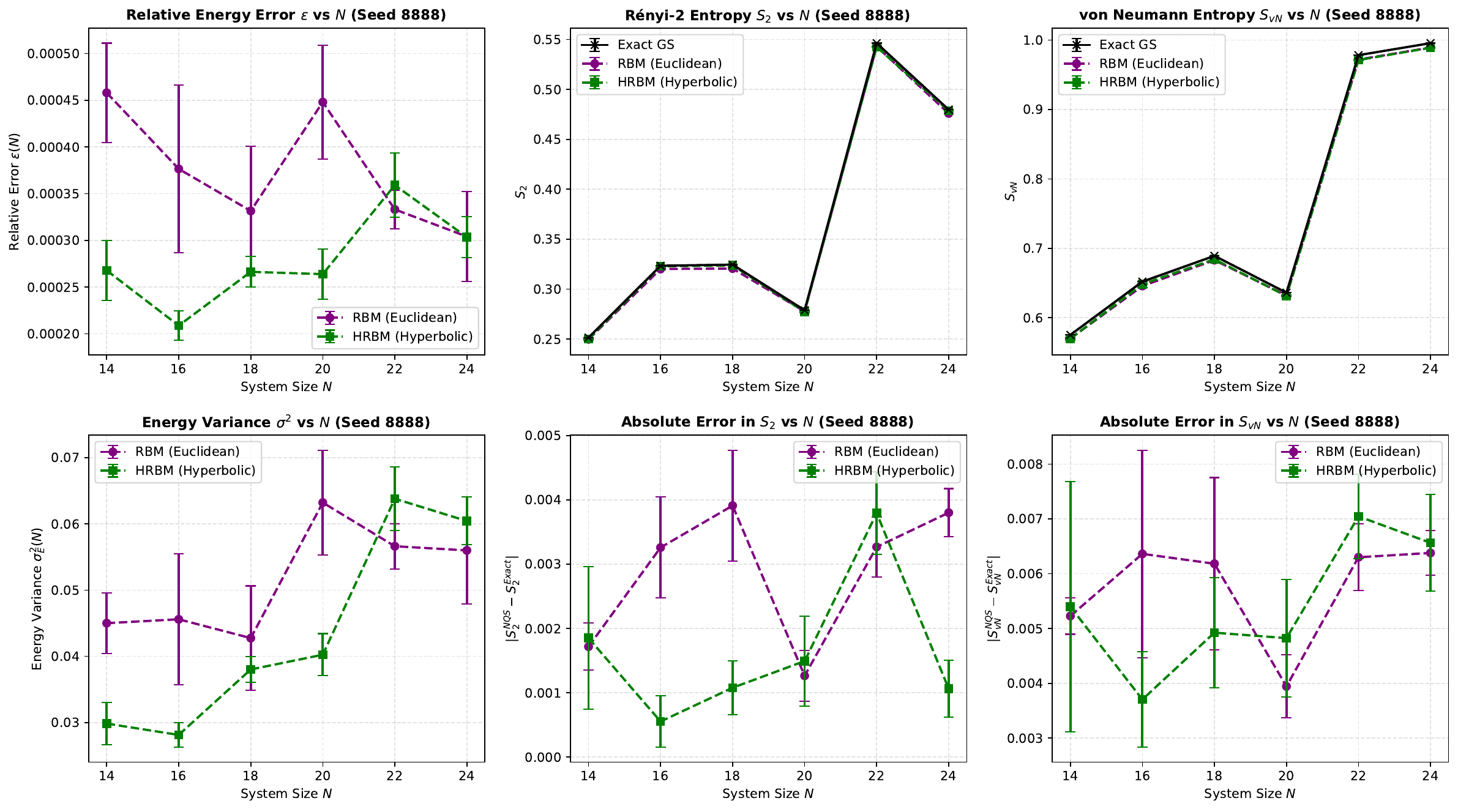}
\caption{Scaling benchmark (QSK seed=8888) for RBM and HRBM in the QSK setting as the system size $N$ increases from $N=14$ to $N=24$. The error bars are obtained by averaging over different NQS random seeds.}\label{scaling_s8888}
\end{figure}
\FloatBarrier

\begin{figure}[!ht]
\centering
\includegraphics[width=.9\textwidth]{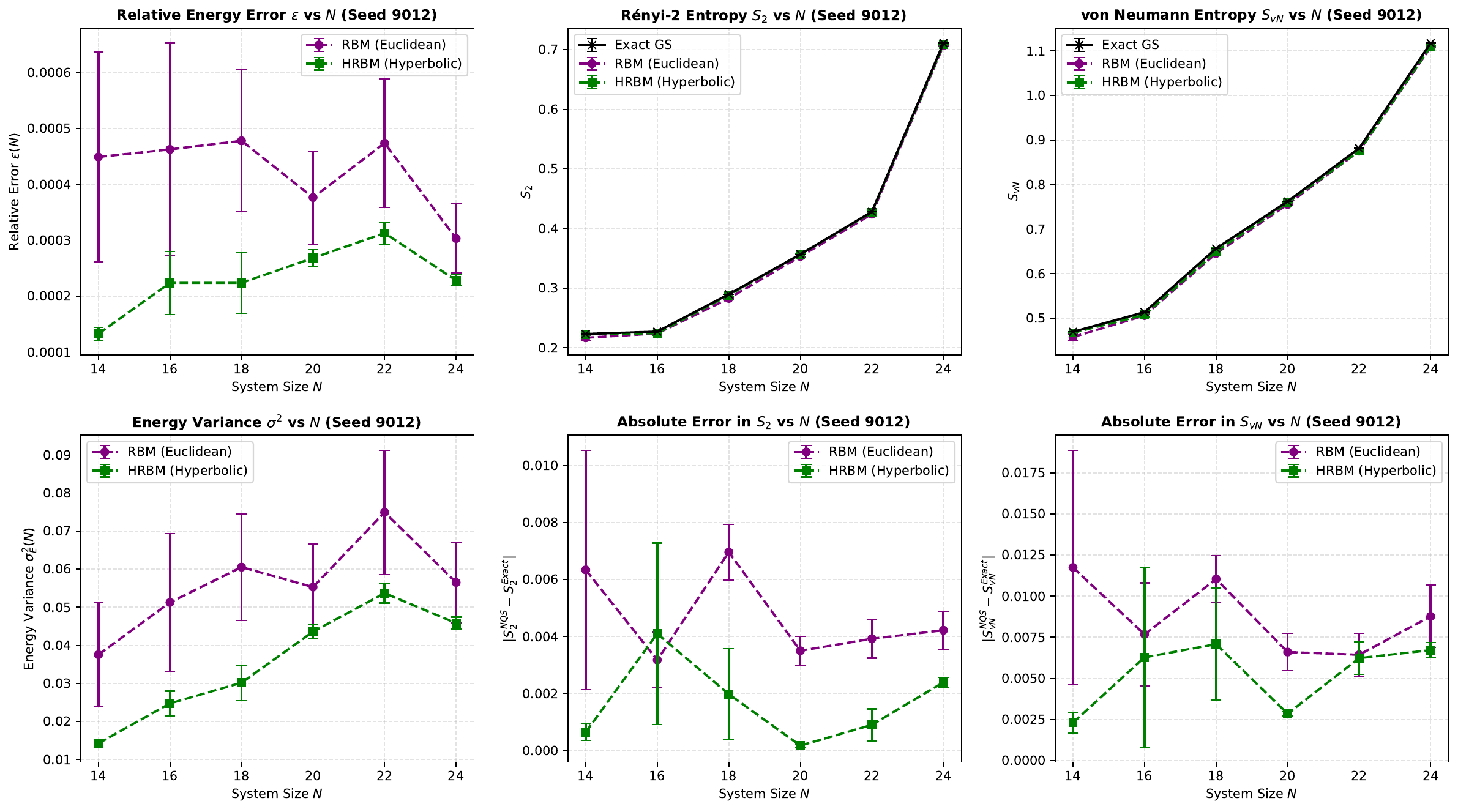}
\caption{Scaling benchmark (QSK seed=9012) for RBM and HRBM in the QSK setting as the system size $N$ increases from $N=14$ to $N=24$. The error bars are obtained by averaging over different NQS random seeds.}\label{scaling_s9012}
\end{figure}
\FloatBarrier
\begin{figure}[!ht]
\centering
\includegraphics[width=.9\textwidth]{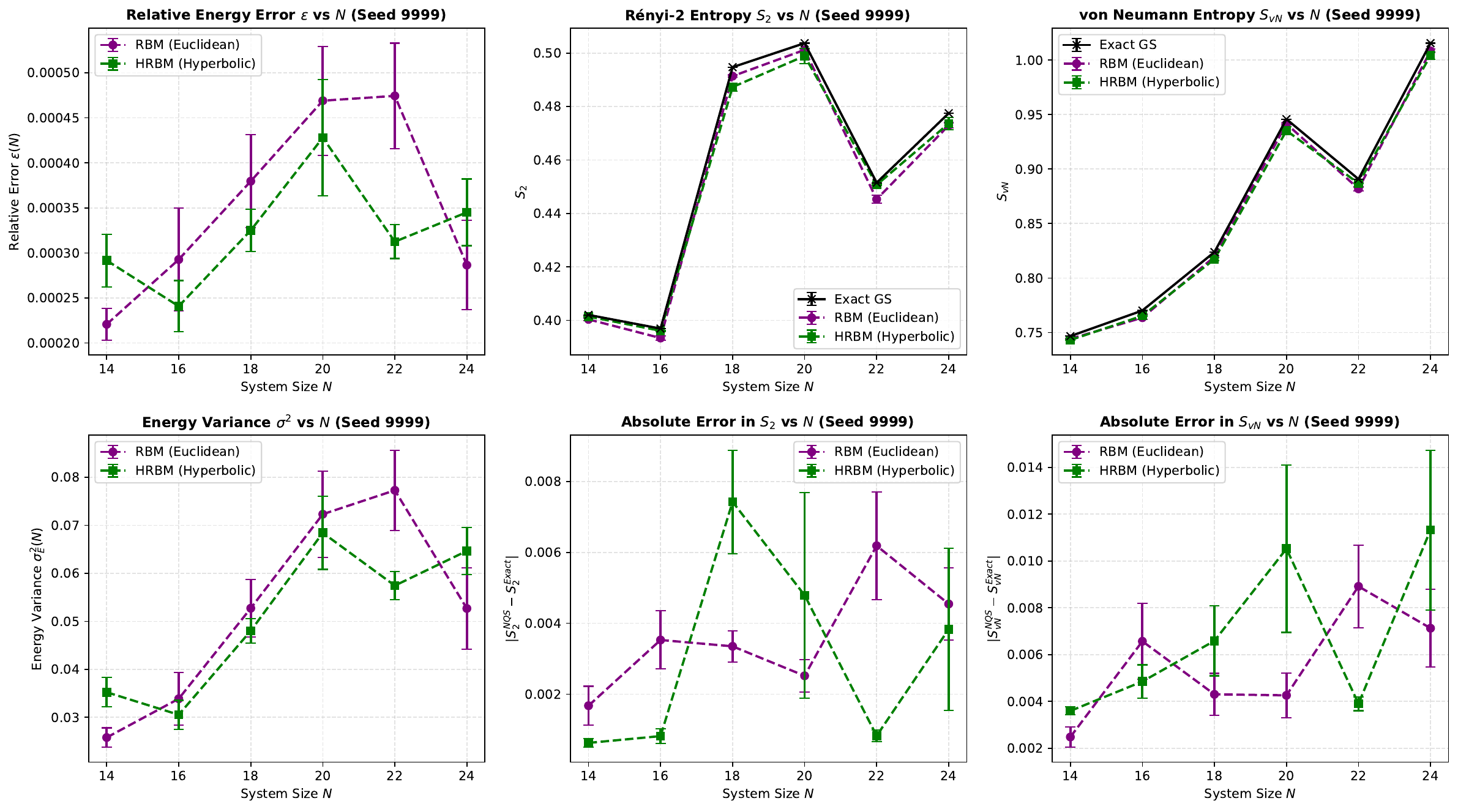}
\caption{Scaling benchmark (QSK seed=9999) for RBM and HRBM in the QSK setting as the system size $N$ increases from $N=14$ to $N=24$. The error bars are obtained by averaging over different NQS random seeds.}\label{scaling_s9999}
\end{figure}
\FloatBarrier
\clearpage
\subsection{Full entanglement spectra at $N=14$ to $N=22$} \label{ent_plots}
\begin{figure}[!ht]
\centering
\includegraphics[width=.9\textwidth]{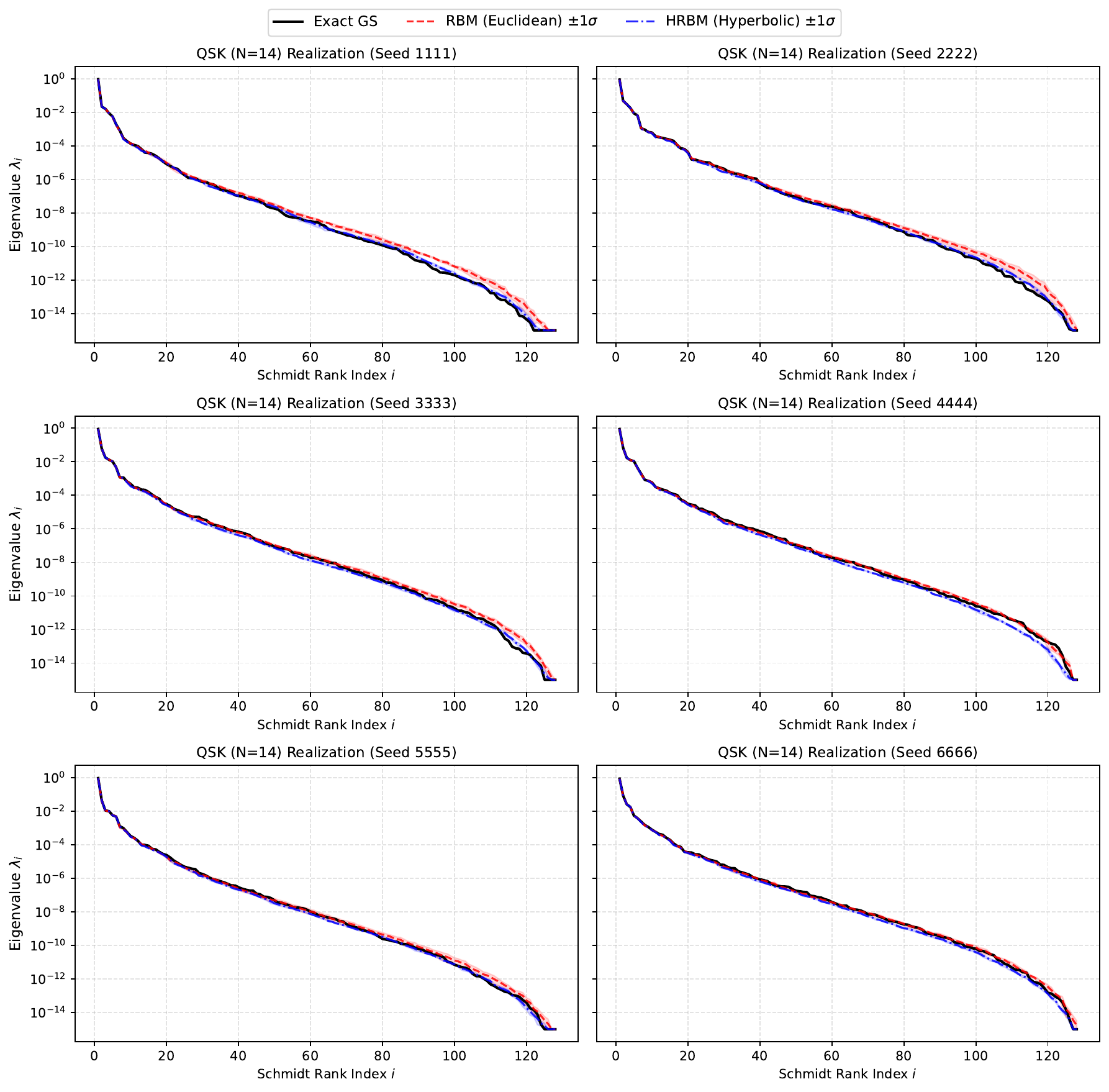}
\caption{Full entanglement spectra of RBM and HRBM NQS in QSK VMC setting with $N=14$ at different QSK model random seeds (1111, 2222, 3333, 4444, 5555, 6666). In each subplot, the RBM and HRBM lines denote the mean value obtained by averaging over different VMC runs involving different NQS random seeds, with shading indicating one standard deviation.}\label{ent_spec_N14_1}
\end{figure}
\FloatBarrier

\begin{figure}[!ht]
\centering
\includegraphics[width=.9\textwidth]{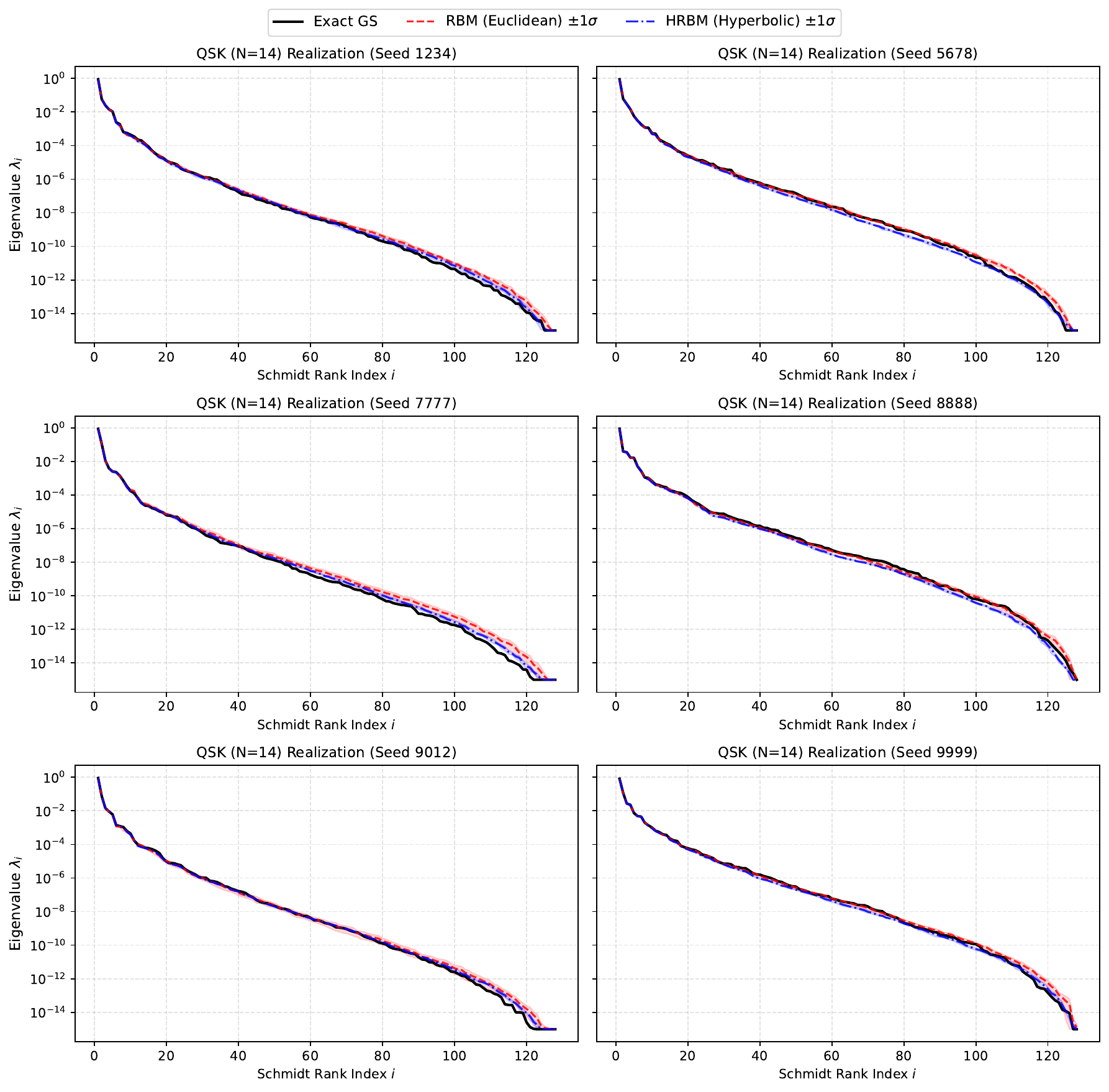}
\caption{Full entanglement spectra of RBM and HRBM NQS in QSK VMC setting with $N=14$ at different QSK model random seeds (1234, 5678, 7777, 8888, 9012, 9999). In each subplot, the RBM and HRBM lines denote the mean value obtained by averaging over different VMC runs involving different NQS random seeds, with shading indicating one standard deviation.}\label{ent_spec_N14_2}
\end{figure}
\FloatBarrier
\begin{figure}[!ht]
\centering
\includegraphics[width=.9\textwidth]{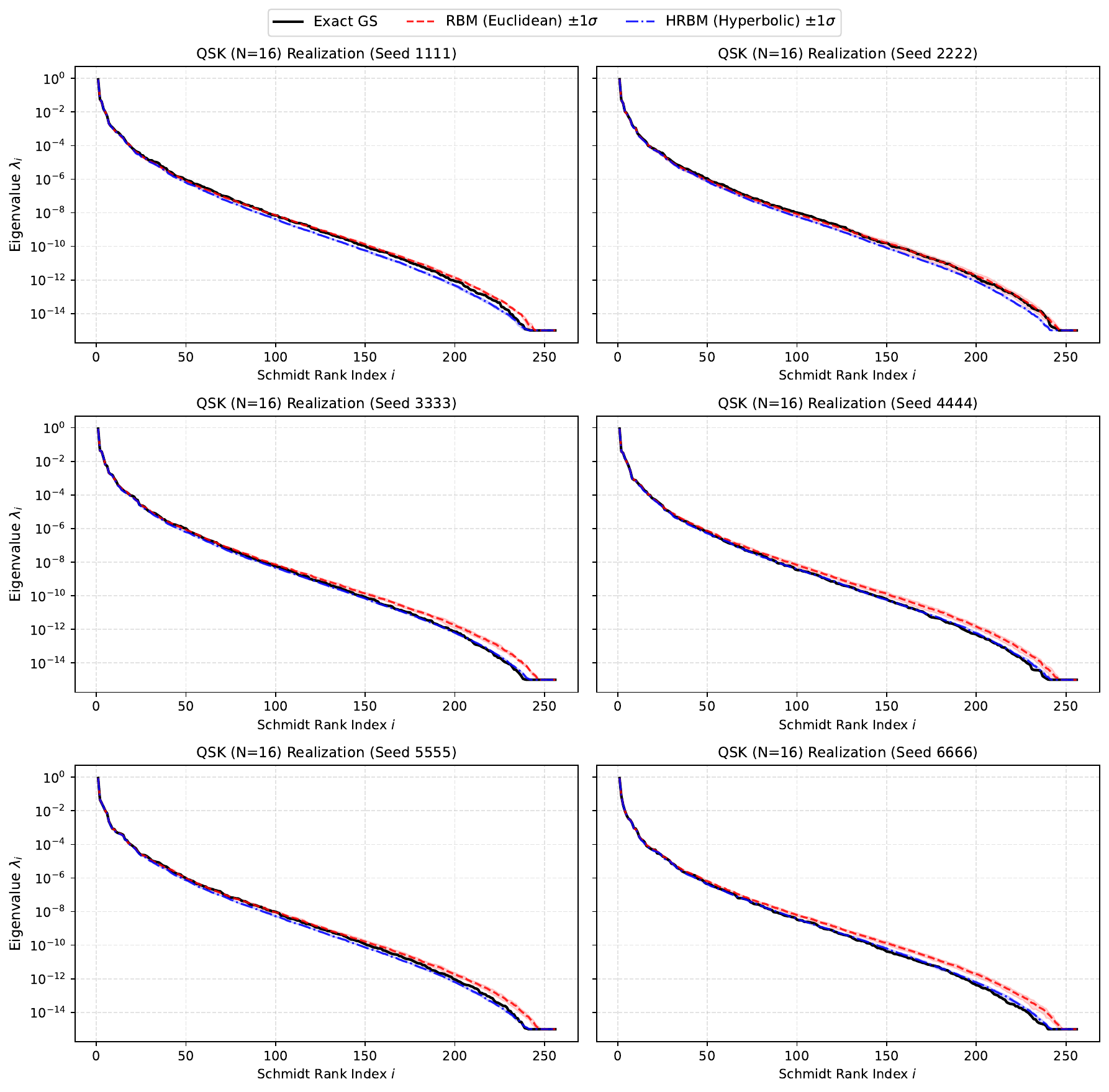}
\caption{Full entanglement spectra of RBM and HRBM NQS in QSK VMC setting with $N=16$ at different QSK model random seeds (1111, 2222, 3333, 4444, 5555, 6666). In each subplot, the RBM and HRBM lines denote the mean value obtained by averaging over different VMC runs involving different NQS random seeds, with shading indicating one standard deviation.}\label{ent_spec_N16_1}
\end{figure}
\FloatBarrier

\begin{figure}[!ht]
\centering
\includegraphics[width=.9\textwidth]{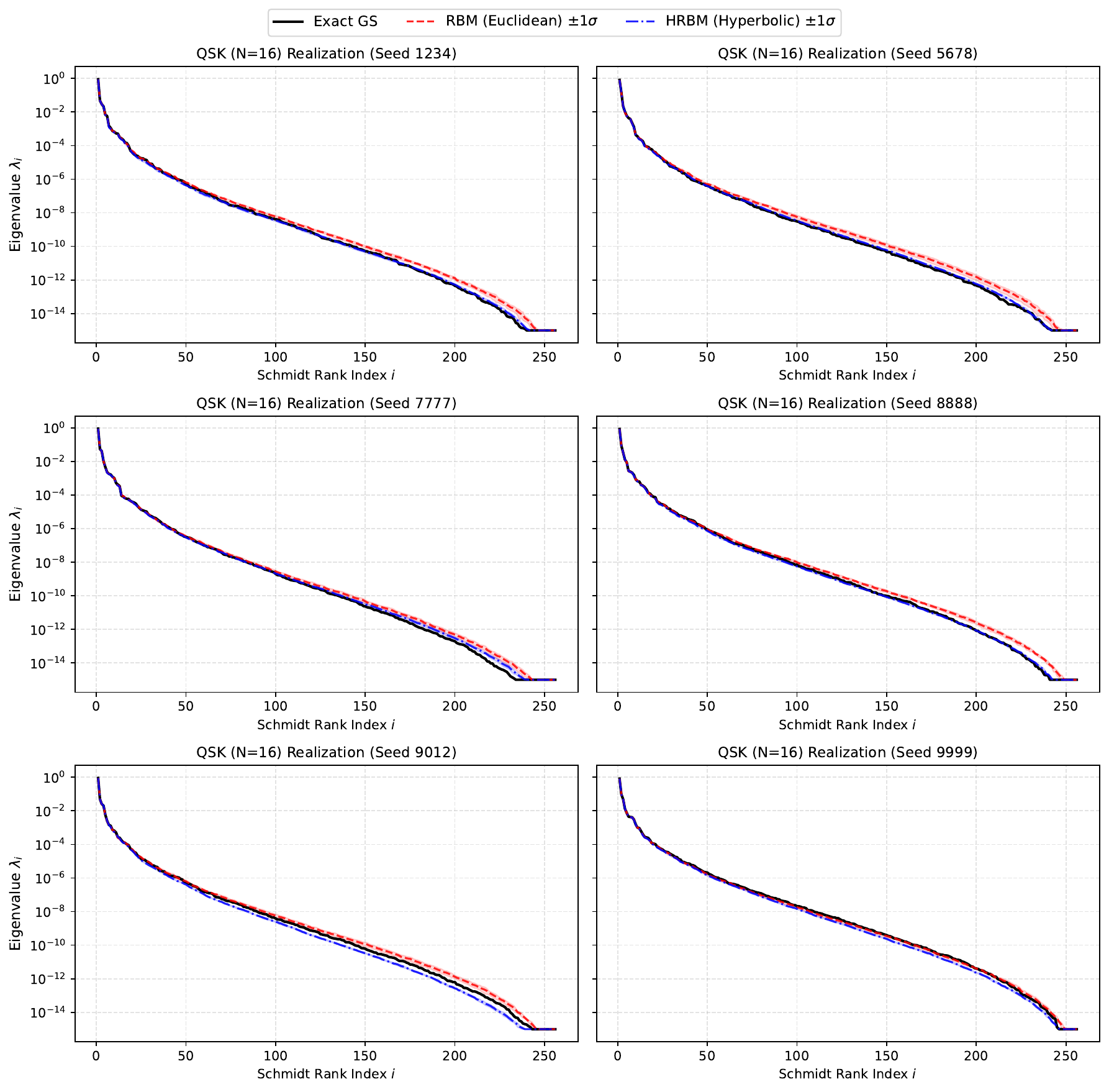}
\caption{Full entanglement spectra of RBM and HRBM NQS in QSK VMC setting with $N=16$ at different QSK model random seeds (1234, 5678, 7777, 8888, 9012, 9999). In each subplot, the RBM and HRBM lines denote the mean value obtained by averaging over different VMC runs involving different NQS random seeds, with shading indicating one standard deviation.}\label{ent_spec_N16_2}
\end{figure}
\FloatBarrier

\begin{figure}[!ht]
\centering
\includegraphics[width=.9\textwidth]{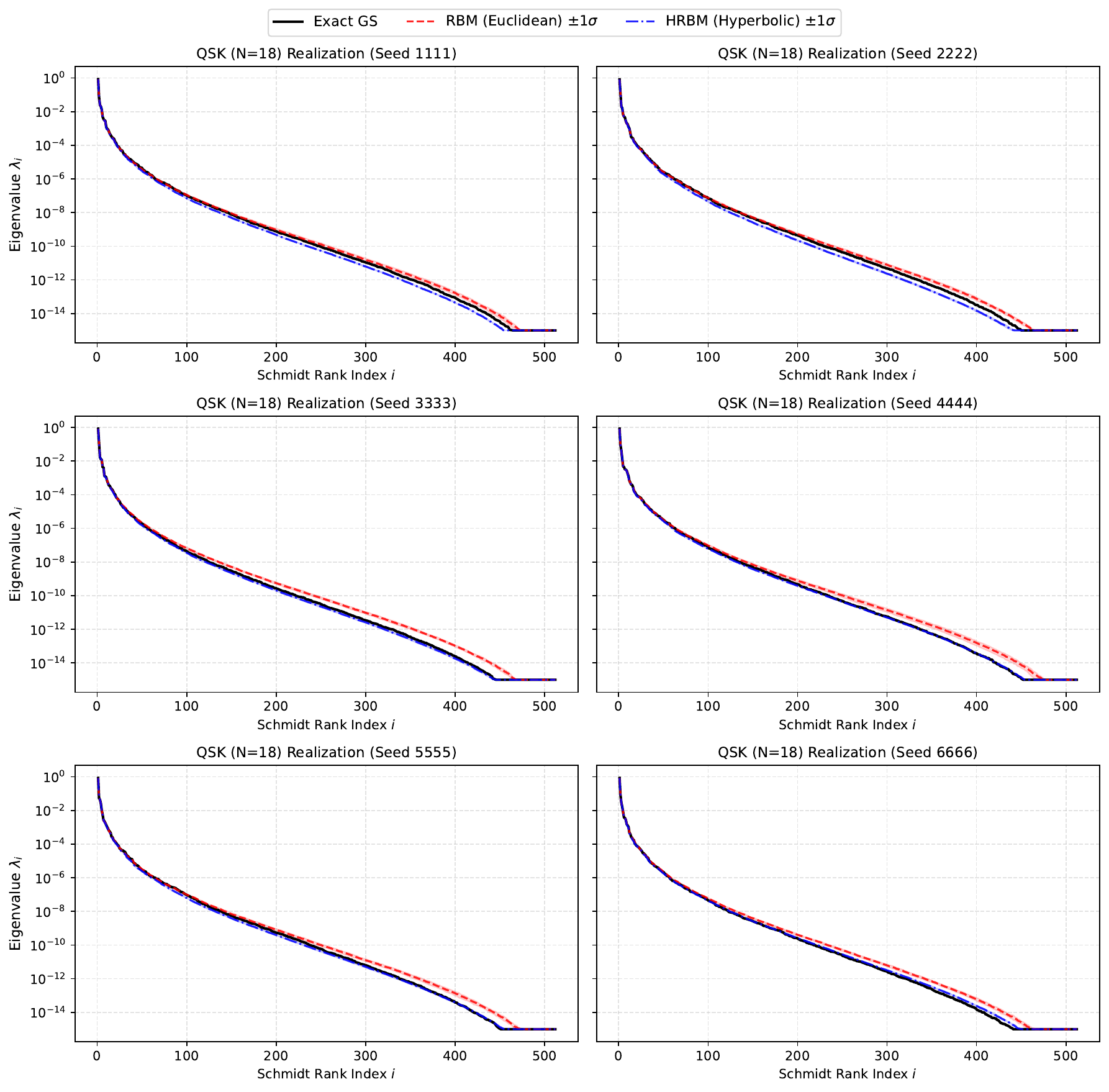}
\caption{Full entanglement spectra of RBM and HRBM NQS in QSK VMC setting with $N=18$ at different QSK model random seeds (1111, 2222, 3333, 4444, 5555, 6666). In each subplot, the RBM and HRBM lines denote the mean value obtained by averaging over different VMC runs involving different NQS random seeds, with shading indicating one standard deviation.}\label{ent_spec_N18_1}
\end{figure}
\FloatBarrier

\begin{figure}[!ht]
\centering
\includegraphics[width=.9\textwidth]{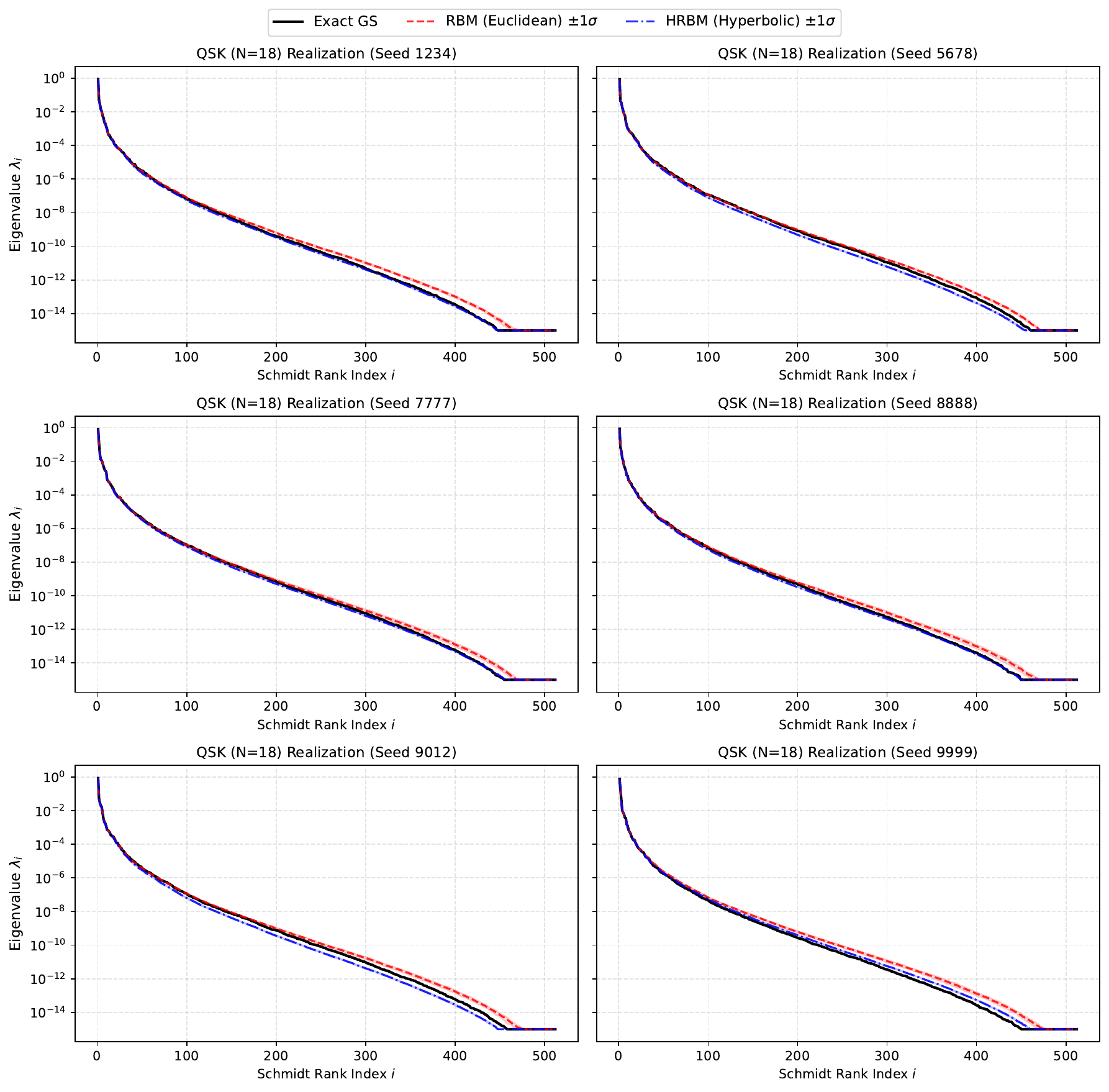}
\caption{Full entanglement spectra of RBM and HRBM NQS in QSK VMC setting with $N=18$ at different QSK model random seeds (1234, 5678, 7777, 8888, 9012, 9999). In each subplot, the RBM and HRBM lines denote the mean value obtained by averaging over different VMC runs involving different NQS random seeds, with shading indicating one standard deviation.}\label{ent_spec_N18_2}
\end{figure}
\FloatBarrier

\begin{figure}[!ht]
\centering
\includegraphics[width=.9\textwidth]{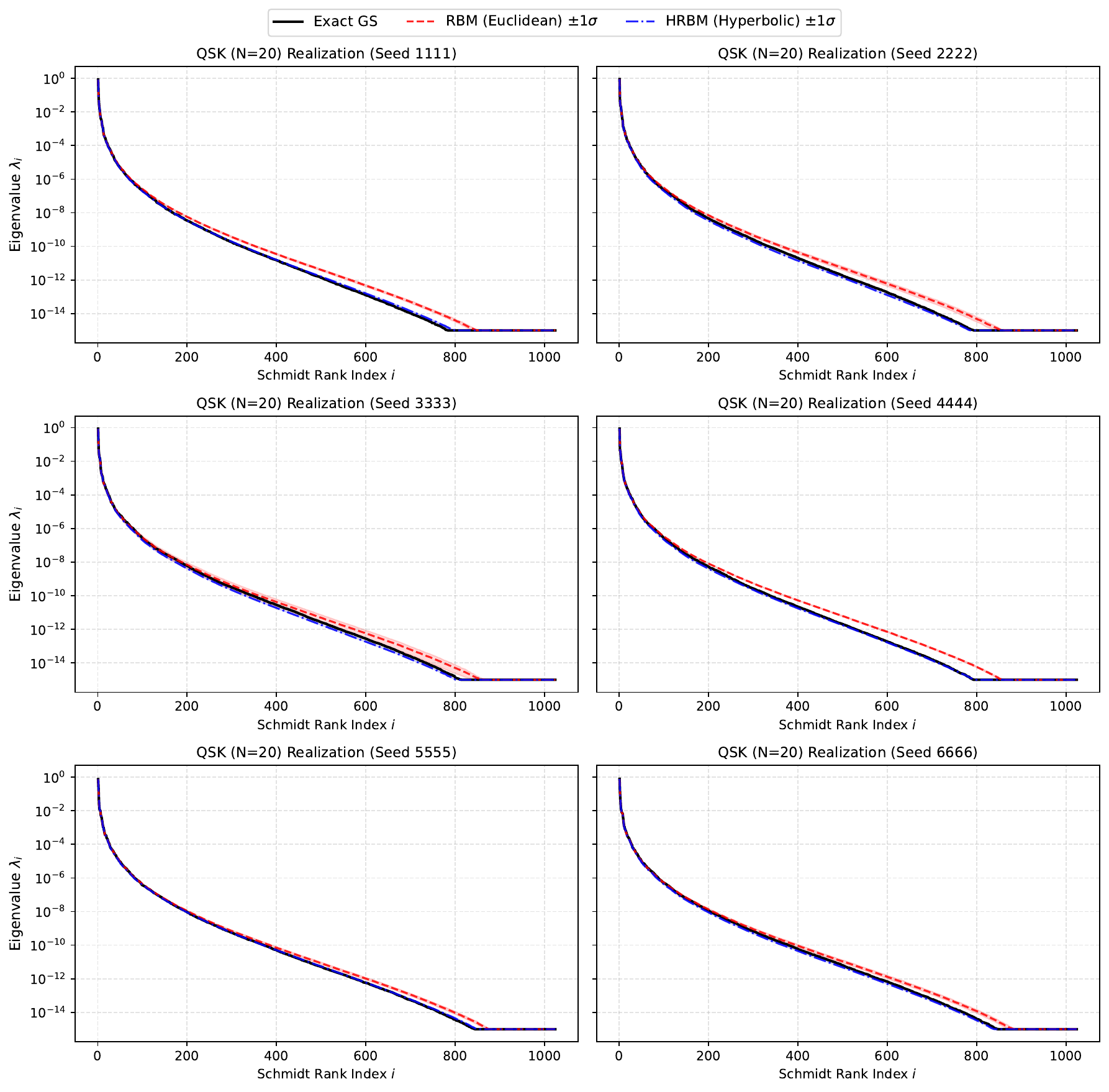}
\caption{Full entanglement spectra of RBM and HRBM NQS in QSK VMC setting with $N=20$ at different QSK model random seeds (1111, 2222, 3333, 4444, 5555, 6666). In each subplot, the RBM and HRBM lines denote the mean value obtained by averaging over different VMC runs involving different NQS random seeds, with shading indicating one standard deviation.}\label{ent_spec_N20_1}
\end{figure}
\FloatBarrier

\begin{figure}[!ht]
\centering
\includegraphics[width=.9\textwidth]{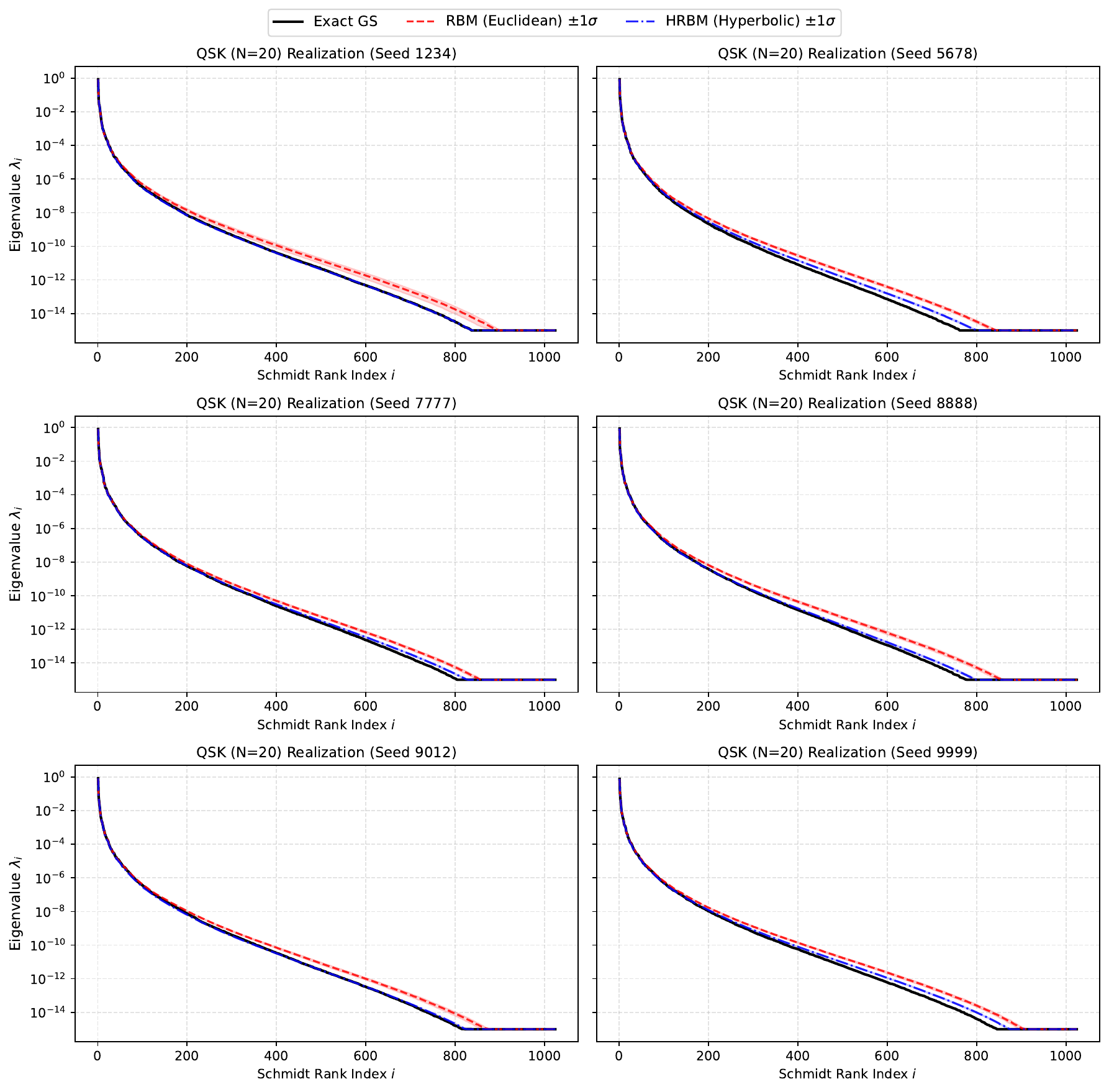}
\caption{Full entanglement spectra of RBM and HRBM NQS in QSK VMC setting with $N=20$ at different QSK model random seeds (1234, 5678, 7777, 8888, 9012, 9999). In each subplot, the RBM and HRBM lines denote the mean value obtained by averaging over different VMC runs involving different NQS random seeds, with shading indicating one standard deviation.}\label{ent_spec_N20_2}
\end{figure}
\FloatBarrier

\begin{figure}[!ht]
\centering
\includegraphics[width=.9\textwidth]{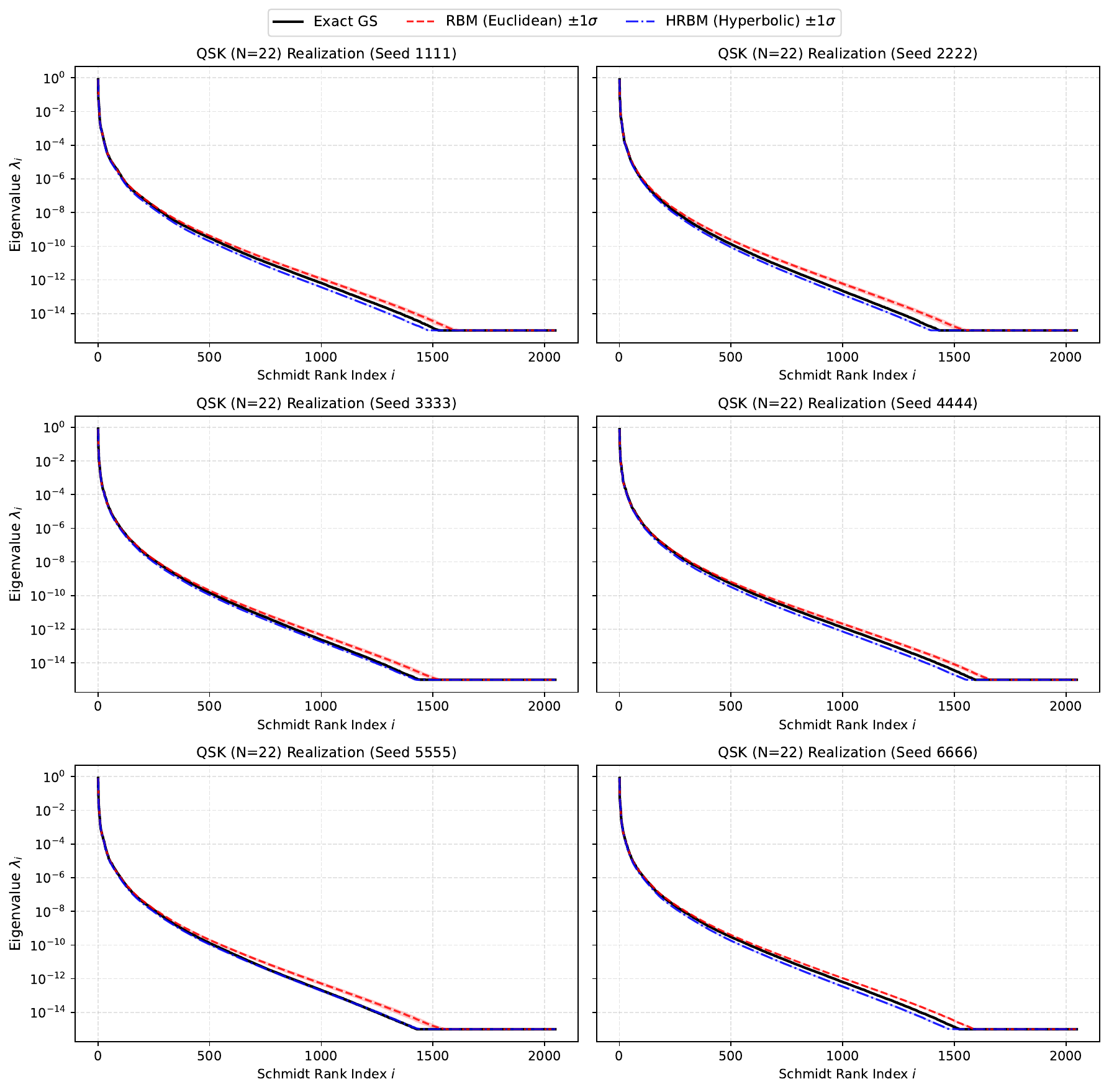}
\caption{Full entanglement spectra of RBM and HRBM NQS in QSK VMC setting with $N=22$ at different QSK model random seeds (1111, 2222, 3333, 4444, 5555, 6666). In each subplot, the RBM and HRBM lines denote the mean value obtained by averaging over different VMC runs involving different NQS random seeds, with shading indicating one standard deviation.}\label{ent_spec_N22_1}
\end{figure}
\FloatBarrier

\begin{figure}[!ht]
\centering
\includegraphics[width=.9\textwidth]{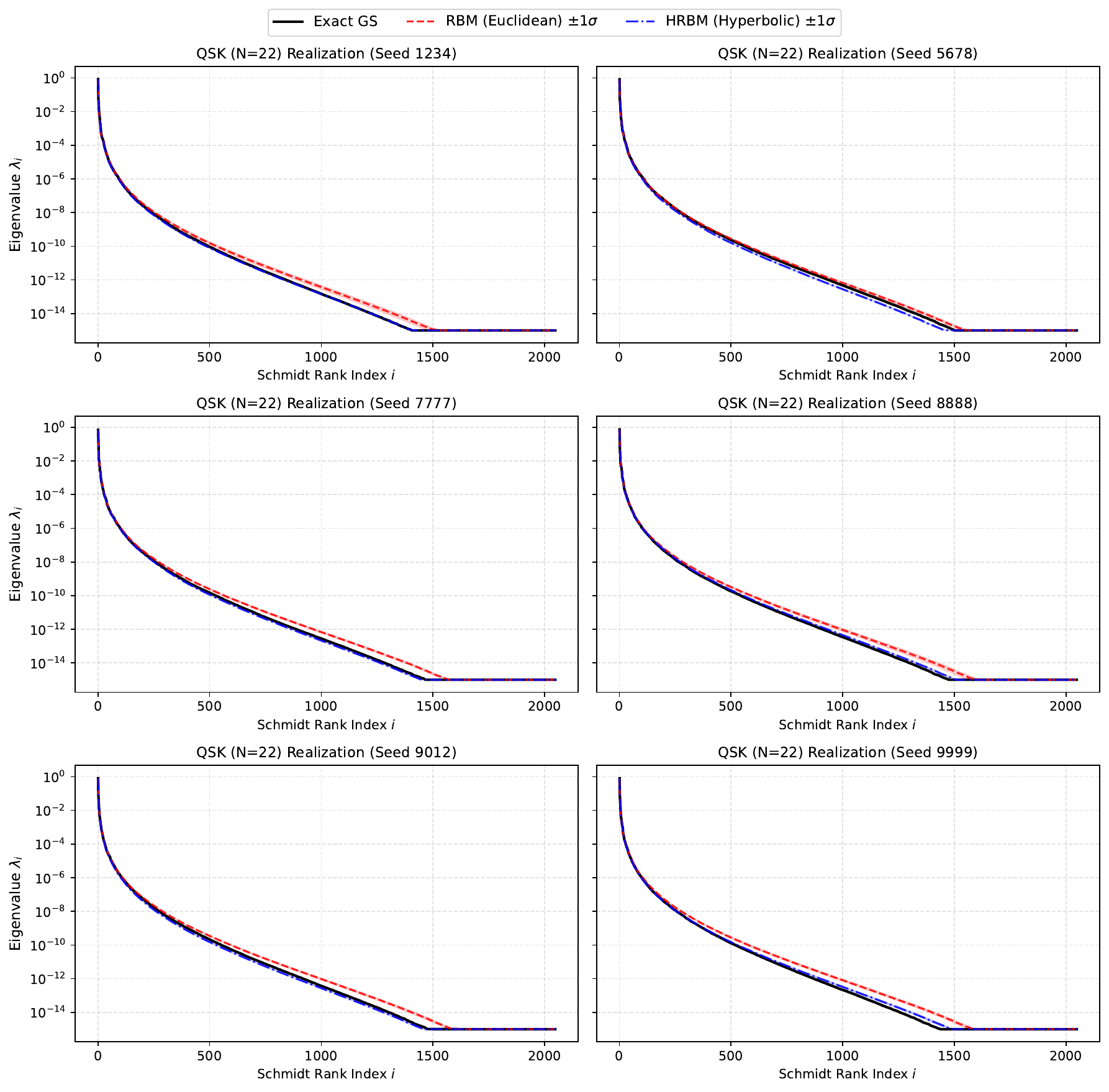}
\caption{Full entanglement spectra of RBM and HRBM NQS in QSK VMC setting with $N=22$ at different QSK model random seeds (1234, 5678, 7777, 8888, 9012, 9999). In each subplot, the RBM and HRBM lines denote the mean value obtained by averaging over different VMC runs involving different NQS random seeds, with shading indicating one standard deviation.}\label{ent_spec_N22_2}
\end{figure}
\FloatBarrier
\clearpage
\subsection{Effects of the spatial constraint hyperparameter $L_{max}$} \label{lmax_effects}
\begin{figure}[!ht]
\centering
\includegraphics[width=.8\textwidth]{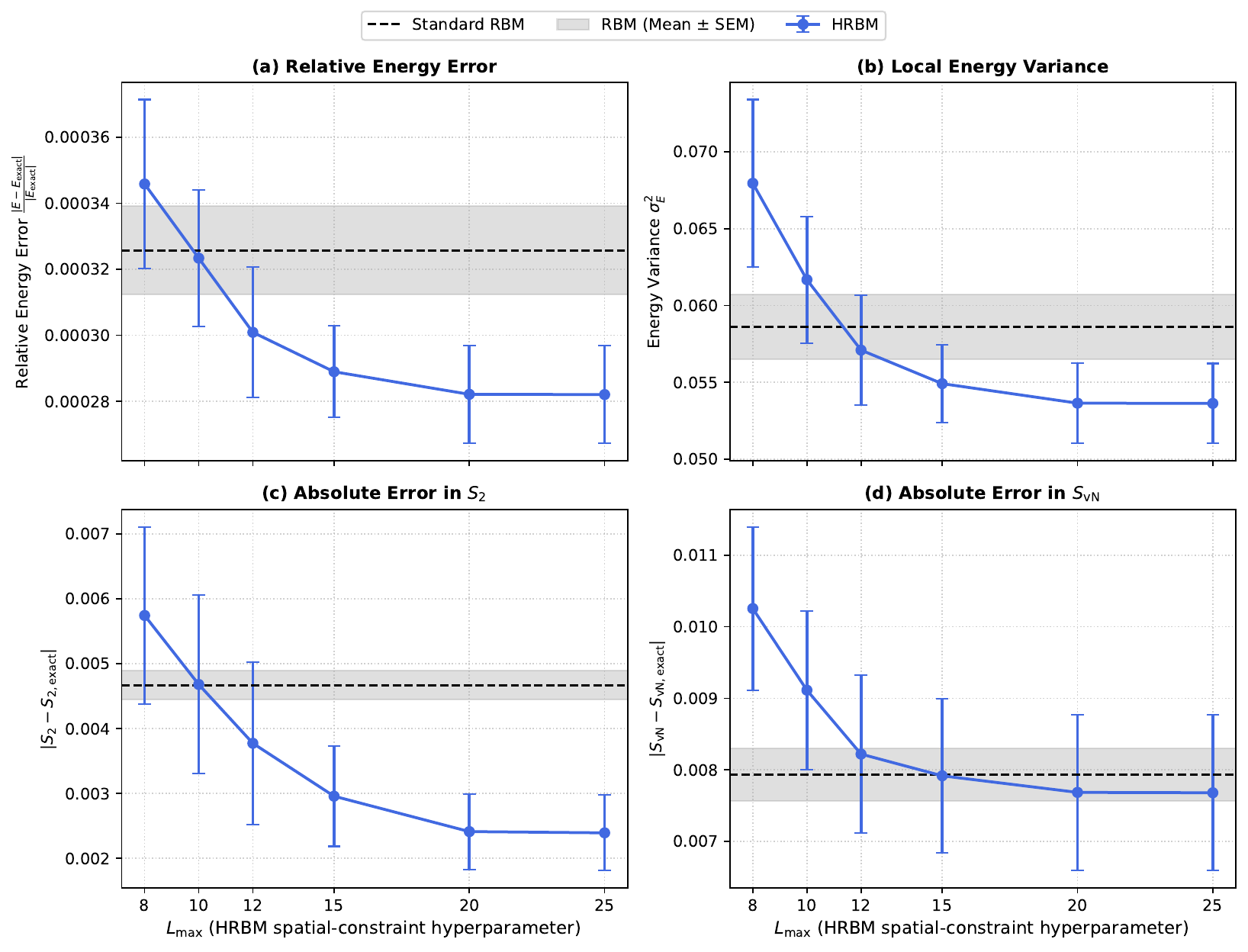}
\caption{Effects of the spatial constraint hyperparameter $L_{max}$ on the performance of HRBM at $N=24$ on (a) the relative energy error $\varepsilon$ (top left subfigure), (b) local energy variance $\sigma(\varepsilon)$ (top right subfigure), (c) absolute error in the R\'enyi-2 $S_2$ entropy (bottom left subfigure), and (d) absolute error in von Neumann entropy $S_{vN}$ (bottom right subfigure).
The error bars denote the standard errors obtained by averaging over the 12 different QSK disorder realizations. The black dashed horizontal line with shading denotes the RBM value and its standard error. The data used to plot the subfigures (a) and (b) are listed in Table \ref{effects_lmax_energy}, while the data used to plot the subfigures (c) and (d) are listed in Table \ref{effects_lmax_entropy}.
 As described in the main text, for $N\leq 22$, choosing $L_{max}=10$ suffices to ensure that HRBM outperform RBM but for $N=24$, choosing $L_{max}=10$ leads to an underperformance of HRBM compared to RBM. Enlarging $L_{max}$ to 20 leads to a significant improve in HRBM's performance. While $L_{max}=25$ is marginally better than $L_{max}=20$ in terms of the key performance metrics, $L_{max}=20$ is sufficient for our computational purposes. }\label{n24_lmax_effects}
\end{figure}
\FloatBarrier
\begin{table}[!ht]
\centering
\begin{tabular}{ccccc}
\hline\hline
NQS model  &  Mean relative energy error $\varepsilon$ &   SE$(\varepsilon)$  & Mean variance & SE (variance)\\
\hline\hline
RBM & $3.2570 \times 10^{-4}$ & $1.3413 \times 10^{-5}$ & $5.8616 \times 10^{-2}$ & $2.0967 \times 10^{-3}$ \\
HRBM ($L_{max}=8$) & $3.4581 \times 10^{-4}$ & $2.5561 \times 10^{-5}$ & $6.7955 \times 10^{-2}$ & $5.4445 \times 10^{-3}$ \\
HRBM ($L_{max}=10$) & $3.2336 \times 10^{-4}$ & $2.0718 \times 10^{-5}$ & $6.1678 \times 10^{-2}$ & $4.1216 \times 10^{-3}$ \\
HRBM ($L_{max}=12$) & $3.0091 \times 10^{-4}$ & $1.9719 \times 10^{-5}$ & $5.7099 \times 10^{-2}$ & $3.5536 \times 10^{-3}$ \\
HRBM ($L_{max}=15$) & $2.8898 \times 10^{-4}$ & $1.3856 \times 10^{-5}$ & $5.4916 \times 10^{-2}$ & $2.5220 \times 10^{-3}$ \\
HRBM ($L_{max}=20$) & $2.8214 \times 10^{-4}$ & $1.4744 \times 10^{-5}$ & $5.3648 \times 10^{-2}$ & $2.5943 \times 10^{-3}$ \\
HRBM ($L_{max}=25$) & $2.8206 \times 10^{-4}$ & $1.4768 \times 10^{-5}$ & $5.3631 \times 10^{-2}$ & $2.5989 \times 10^{-3}$ \\
\hline
\end{tabular}
\caption{This table lists the mean relative energy error $\varepsilon$ and its associated standard error (SE), as well as mean variance and its associated standard error of HRBM NQS at different $L_{max}$ and RBM NQS. These values are are obtained by averaging over different QSK realizations (as well as NQS initialization seeds).} \label{effects_lmax_energy}
\end{table}

\begin{table}[!ht]
\centering
\begin{tabular}{ccccc}
\hline\hline
NQS & Mean $|S_2^\text{err}|$ &   SE($|S_2^\text{err}|$) &  Mean $|S_{vN}^\text{err}|$ &  SE($|S_{vN}^\text{err}|$) \\
\hline\hline
RBM & $4.6705 \times 10^{-3}$ & $2.1982 \times 10^{-4}$ & $7.9337 \times 10^{-3}$ & $3.6607 \times 10^{-4}$ \\
HRBM ($L_{max}=8$) & $5.7437 \times 10^{-3}$ & $1.3664 \times 10^{-3}$ & $1.0255 \times 10^{-2}$ & $1.1421 \times 10^{-3}$ \\
HRBM ($L_{max}=10$) & $4.6798 \times 10^{-3}$ & $1.3751 \times 10^{-3}$ & $9.1136 \times 10^{-3}$ & $1.1087 \times 10^{-3}$ \\
HRBM ($L_{max}=12$) & $3.7732 \times 10^{-3}$ & $1.2492 \times 10^{-3}$ & $8.2201 \times 10^{-3}$ & $1.1046 \times 10^{-3}$ \\
HRBM ($L_{max}=15$) & $2.9577 \times 10^{-3}$ & $7.7342 \times 10^{-4}$ & $7.9163 \times 10^{-3}$ & $1.0813 \times 10^{-3}$ \\
HRBM ($L_{max}=20$) & $2.4120 \times 10^{-3}$ & $5.8159 \times 10^{-4}$ & $7.6845 \times 10^{-3}$ & $1.0879 \times 10^{-3}$ \\
HRBM ($L_{max}=25$) & $2.3925 \times 10^{-3}$ & $5.8198 \times 10^{-4}$ & $7.6796 \times 10^{-3}$ & $1.0883 \times 10^{-3}$ \\
\hline
\end{tabular}
\caption{This table lists the absolute entropy errors ($|S_2^\text{err}| = |S^{NQS}_2- S_2^\text{exact}|$, $|S_{vN}^\text{err}|=S^{NQS}_{vN}- S_{vN}^\text{exact}|$)  and their associated standard errors of HRBM NQS at different $L_{max}$ and RBM NQS. These values are obtained by averaging over 12 different QSK disorder realizations (as well as NQS initialization seeds). } \label{effects_lmax_entropy}
\end{table}
\newpage


\begin{thebibliography}{9}
\addcontentsline{toc}{section}{References}
\bibitem{volume-2017}  D.-L.Deng, X. Li, \&  S. D. Sarma, Quantum entanglement in neural network states. Phys. Rev. X 7, 021021 (2017).

\bibitem{volume-2019}  Y. Levine, O. Sharir , N. Cohen \& A. Shashua, Quantum entanglement in deep learning architectures. Phys. Rev. Lett. 122, 065301 (2019).

\bibitem{nn-volume-1} G. Passetti, D. Hofmann, P. Neitemeier, L. Grunwald, M. A. Sentef, and D. M. Kennes, Can Neural Quantum States Learn Volume-Law Ground States?, Physical Review Letters 131, 036502 (2023).

\bibitem{nn-volume-2} Z. Denis, A. Sinibaldi, and G. Carleo, Comment on 'Can Neural Quantum States Learn Volume-Law Ground States, arXiv:2309.11534v2 [quant-ph]

\bibitem{1606-rbm} G. Carleo and M. Troyer, Solving the Quantum Many-Body Problem with Artificial Neural Networks,  Science 355, 602 (2017), arXiv:1606.02318 [cond-mat.dis-nn]

\bibitem{1610-rbm} L. Huang and L. Wang, Accelerate Monte Carlo Simulations with Restricted Boltzmann Machines, arXiv:1610.02746v2.

\bibitem{1704-ann} Z. Cai and J. Liu, Approximating quantum many-body wave-functions using artificial neural networks,Phys. Rev. B 97, 035116 (2018), arXiv:1704.05148 [cond-mat.str-el]

\bibitem{1709-ann} H. Saito and M. Kato, Machine learning technique to find quantum many-body ground states of bosons on a lattice,   J. Phys. Soc. Jpn. 87, 014001 (2018),  arXiv:1709.05468 [cond-mat.dis-nn]

\bibitem{cnn-1807} X. Liang, Wen-Yuan Liu, Pei-Ze Lin, Guang-Can Guo, Yong-Sheng Zhang, and Lixin He, Solving frustrated quantum many-particle models with convolutional neural networks, arXiv: 1807.09422v2

\bibitem{nqs-2020} T. Westerhout, N. Astrakhantsev, K. S. Tikhonov, M.I . Katsnelson, A. A. Bagrov, Generalization properties of neural network approximations to frustrated magnet ground states, Nature communications 11 (1), 1593

\bibitem{gcnn-2211} C. Roth, A. Szab\'o, and A. H. MacDonald, High-accuracy variational Monte Carlo for frustrated magnets with deep neural networks, Phys. Rev. B 108, 054410 (2023), arXiv:2211.07749v2 [cond-mat.str-el]

\bibitem{rnn_20} M. Hibat-Allah,  M. Ganahl, L. E. Hayward, R. G. Melko, and J. Carrasquill, Recurrent neural network wave functions, Physical Review Research 2, 023358 (2020).

\bibitem{rnn-24} M. Hibat-Allah, E. Merali, G. Torlai, R. G. Melko and J. Carrasquilla, Recurrent neural network wave functions for Rydberg atom arrays on kagome lattice, arXiv:2405.20384v1 [cond-mat.quant-gas]

\bibitem{2306-transformer} K. Sprague and S. Czischek, Variational Monte Carlo with Large Patched Transformers, Commun Phys 7, 90 (2024), arXiv:2306.03921 [quant-ph]

\bibitem{2406-transformer} H. Lange, G. Bornet, G. Emperauger, C. Chen, T. Lahaye, S. Kienle, A. Browaeys, A. Bohrdt, Transformer neural networks and quantum simulators: a hybrid approach for simulating strongly correlated systems,  Quantum 9, 1675 (2025), arXiv:2406.00091 [cond-mat.dis-nn]

\bibitem{vmc_textbook} F. Becca and S. Sorella, Quantum Monte Carlo Approaches for Correlated Systems (Cambridge University Press, 2017)

\bibitem{hnqs-1} H. L. Dao, Hyperbolic recurrent neural network as the first type of non-Euclidean neural quantum state ansatz, Eur. Phys. J. Plus 141:199, arXiv:2505.22083 [quant-ph, cond-mat.dis-nn, cs.LG, physics.comp-ph] (2026)

\bibitem{hnqs-2} H. L. Dao, New non-Euclidean neural quantum states from hyperbolic Lorentz recurrent architectures, arXiv:2604.2337 [quant-ph, cs.LG, cond-mat.dis-nn]

\bibitem{hnqs-3}  H. L. Dao, Two-dimensional Hyperbolic RNN Neural Quantum State, arXiv:2606.25600 [quant-ph]

 \bibitem{ganea-18} O.-E. Ganea, G. Becigneul, and T. Hofmann, Hyperbolic Neural Networks, Advances in Neural Information
Processing Systems 31, pages 5345–5355. Curran Associates, Inc. arXiv: 1805.09112 [cs.LG]

\bibitem{hyp-rep} M. Nickel and D. Kiela, Learning Continuous Hierarchies in the Lorentz Model of Hyperbolic Geometry, ICML 2018, arXiv:1806.03417 [cs.AI]
\bibitem{hypnn-21}
 W. Chen, X. Han, Y. Lin, H. Zhao, Z. Liu, P. Li, M. Sun, J. Zhou, Fully Hyperbolic Neural Networks, in ACL 2022 Main Conference, arXiv:2105.14686 [cs.CL].

\bibitem{hyp-survey} W. Peng, T. Varanka, A. Mostafa, H. Shi, G. Zhao, Hyperbolic Deep Neural Networks: A Survey, arXiv:2101.04562 [cs.LG, cs.CV]

\bibitem{hyp-graph-19} Q. Liu, M. Nickel and D. Kiela, Hyperbolic Graph Neural Network, arXiv:1910.12892 [cs.LG]

\bibitem{hyp-graph-chami} I. Chami, R. Ying, C. Re, and J. Leskovec, Hyperbolic Graph Convolutional Neural Networks, arXiv: 1910.12933 [cs.LG]  

\bibitem{sarkar} R. Sarkar, Low distortion Delaunay embedding of trees in hyperbolic plane, In Proceeding of the International Symposium on Graph Drawing (GD 2011), (Eindhoven, Netherlands, 2011), p. 355–366

\bibitem{qsk-ref-1} D. Sherrington and S. Kirkpatrick, Solvable Model of a Spin-Glass, Physical Review Letters 35, 1792 (1975).

\bibitem{qsk-ref-2}
P. M. Schindler, T. Guaita, T. Shi, E. Demler, and J. I. Cirac, A Variational Ansatz for the Ground State of the Quantum Sherrington-Kirkpatrick Model, arXiv:2204.02923v2

\bibitem{netket-1} G. Carleo, K. Choo, D. Hofmann, J. Smith, T. Westerhout, F. Alet, E. J. Davis, S. Efthymiou, I. Glasser, S.-H. Lin, M. Mauri, G. Mazzola, C. B. Mendl, E. van Nieuwenburg, O. O'reilly, H. Th\'eveniaut, G. Torlai, F. Vicentini, A. Wietek, NetKet: A machine learning toolkit for many-body quantum systems, SoftwareX, 10, 100311 (2019)

\bibitem{netket-3} F. Vicentini, D. Hofmann, A. Szabo, D. Wu, C. Roth, C. Giuliani, G. Pescia, J. Nys, V. Vargas-Calderon, N. Astrakhantsev, and G. Carleo, NetKet 3: Machine Learning Toolbox for Many-Body Quantum Systems, SciPost Phys. Codebases, 7 (2022).

\bibitem{jax2018github}
J. Bradbury, R. Frostig, P. Hawkins, M. J. Johnson, et. al., JAX: composable transformations of {Python}+{NumPy} programs (version 0.3.13), 2018, \href{http://github.com/jax-ml/jax}{http://github.com/jax-ml/jax}

\bibitem{flax2020github}
J. Heek, A. Levskaya, A. Oliver, M.Ritter, et. al, Flax: A neural network library and ecosystem for JAX (version 0.12.9), 2020, \href{http://github.com/google/flax}{http://github.com/google/flax}

\bibitem{ads-cft} J. Maldacena, The large N limit of superconformal field theories and supergravity, Advances in Theoretical and Mathematical Physics. 2 (4): 231–252. arXiv:hep-th/9711200.

\end{thebibliography}
\end{document}